\documentclass[usenatbib,fleqn]{mnras}

\usepackage{newtxtext}
\usepackage{newtxmath}

\usepackage[T1]{fontenc}
\usepackage{ae,aecompl}

\usepackage{graphicx}	
\usepackage{amsmath}	
\usepackage{amssymb}	

\usepackage{times}
\usepackage{xspace}

\newcommand{\wmap}{\textsl{WMAP7}\xspace}
\newcommand{\planck}{\textsl{Planck}\xspace}
\newcommand{\xmm}{\textsl{XMM Newton}\xspace}

\newcommand{\beq}{\begin{equation}}
\newcommand{\eeq}{\end{equation}}
\newcommand{\beqa}{\begin{eqnarray}}
\newcommand{\eeqa}{\end{eqnarray}}

\newcommand{\eg}{e.g.,\xspace}

\def\msun{{\rm M}_{\odot}}
\def\der{{\rm d}}

\title[Probing hot gas around luminous red galaxies]{Probing hot gas around luminous red galaxies through the Sunyaev-Zel'dovich effect}

\author[Tanimura et al.]{Hideki Tanimura,$^{1}$\thanks{E-mail: hideki.tanimura@ias.u-psud.fr}
Gary Hinshaw,$^{2,3,4}$
Ian G. McCarthy,$^{5}$ 
Ludovic Van Waerbeke,$^{2}$\newauthor
Nabila Aghanim,$^{1}$
Yin-Zhe Ma,$^{6,7}$
Alexander Mead,$^{2}$
Tilman Tr\"{o}ster,$^{8}$
Alireza Hojjati$^{2}$ \newauthor
and Bruno Moraes$^{9,10}$
\\
$^{1}$Institut d'Astrophysique Spatiale, CNRS/Universit{\'e} Paris-Sud, Universit{\'e} Paris-Saclay, B{\^a}timent 121, Universit{\'e} Paris-Sud, 91405 Orsay Cedex, France\\
$^{2}$Department of Physics and Astronomy, University of British Columbia, Vancouver, BC V6T 1Z1, Canada\\
$^{3}$Canadian Institute for Advanced Research, 180 Dundas St W, Toronto, ON M5G 1Z8, Canada\\
$^{4}$Canada Research Chair in Observational Cosmology\\
$^{5}$Astrophysics Research Institute, Liverpool John Moores University, Liverpool, L3 5RF, United Kingdom\\
$^{6}$Astrophysics and Cosmology Research Unit, School of Chemistry and Physics,
University of KwaZulu-Natal, Durban, South Africa\\
$^{7}$NAOC-UKZN Computational Astrophysics Center (NUCAC), University of KwaZulu-Natal, Durban, 4000\\
$^{8}$Institute for Astronomy, University of Edinburgh, Royal Observatory, Blackford Hill, Edinburgh, EH9 3HJ, UK\\
$^{9}$Department of Physics and Astronomy, University College London, Gower Street, London, WC1E 6BT, UK\\
$^{10}$Instituto de Fisica, Universidade Federal do Rio de Janeiro, 21941-972, Rio de Janeiro, RJ, Brazil
}

\begin{document}
\label{firstpage}
\pagerange{\pageref{firstpage}--\pageref{lastpage}}
\maketitle

\begin{abstract}
We construct the mean thermal Sunyaev-Zel'dovich (tSZ) Comptonization $y$ profile around Luminous Red Galaxies (LRGs) in the redshift range 0.16 < $z$ < 0.47 from the Sloan Digital Sky Survey (SDSS) Data Release 7 (DR7) using the \planck $y$ map. We detect a significant tSZ signal out to $\sim30$ arcmin, which is well beyond the 10 arcmin angular resolution of the $y$ map and well beyond the virial radii of the LRGs. We compare the measured profile with predictions from the cosmo-OWLS suite of cosmological hydrodynamical simulations. The best agreement is obtained for models that include efficient feedback from active galactic nuclei (AGN), over and above feedback associated with star formation. We also compare our results with predictions based on the halo model with a universal pressure profile (UPP) giving the $y$ signal. The predicted profile is consistent with the data when using stacked weak lensing measurements to estimate the halo masses of the LRGs, but only if we account for the clustering of neighbouring haloes via a two-halo term.
\end{abstract}




%
%

\section{Introduction}

In the standard $\Lambda$CDM cosmological paradigm more than $95\%$ of the energy density in the Universe is in the form of dark matter and dark energy, whereas baryonic matter only comprises $\simeq5\%$ \citep{Hinshaw2013, Planck2016-I}. While the evolution of the homogeneous Universe and of small density perturbations is well understood, the details of the complicated structure-formation process that results in the observed distribution and properties of galaxies are more elusive. The general picture is that galaxies form at the knots of a dark-matter skeleton, but the details of how gas is converted into stars, and how the electro-magnetic spectrum of a galaxy arises, are not well understood. One important tracer of cosmological structure are clusters of galaxies, which are the most massive bound structures and which mark prominent density peaks of the large-scale structure. The distribution and properties of galaxy clusters are therefore powerful tools for understanding both cosmological structure formation and galaxy evolution.

X-ray observations of clusters have discovered that they are intense sources of high-energy radiation that is emitted by a hot gas ($T \sim 10^7$ K) located between member galaxies. This intergalactic gas (or intracluster medium, ICM) contains significantly more baryons than are contained in all the stars in the galaxies and indicates a complex dynamical evolution of the ICM regulated by the radiative cooling and non-gravitational heating from stellar sources and, particularly, active galactic nuclei (AGN). AGN feedback has a wide range of impacts on galaxies and galaxy clusters: the observed relation between the central super-massive black hole mass and stellar bulge velocity dispersion, the regulation of cool cores, and the suppression of star formation in massive galaxies predicted by N-body simulations (e.g., \citealt{Schneider2006, Gitti2012}). Thus, the interplay of hot gas with the relativistic plasma ejected by the AGN is key for understanding the growth and evolution of galaxies and the formation of large-scale structure. It has become clear that AGN feedback effects on the ICM must be incorporated in any model of galaxy evolution (e.g., \citealt{Sijacki2007, Battaglia2010, Schaye2010, Vogelsberger2013, McCarthy2014, Steinborn2015}). However, non-gravitational processes such as gas dynamics, heating and radiative cooling are not well understood. If one is interested in studying the effect of non-gravitational processes specifically then galaxy groups and low-mass clusters are ideal laboratories since they have shallower gravitational potentials compared to massive clusters and therefore the impact of non-gravitational processes on their formation and evolution may be more noticeable (e.g., \citealt{Johnson2009, Dong2010, Giodini2010, Battaglia2012, Brun2014}).

In addition to X-ray emission, the thermal Sunyaev-Zel'dovich \citep[tSZ;][]{Zeldovich1969, Sunyaev1970, Sunyaev1972, Sunyaev1980} effect provides a way to study hot cluster gas. The tSZ effect arises via a boost to the energy of cool CMB photons as they pass relatively energetic hot electrons and provides an excellent tool for studying the thermodynamic state of the ICM. The tSZ effect is proportional to the pressure of the ICM and therefore has a linear dependence on gas density, compared to a quadratic dependence of X-ray emissivity on density.  This results in a comparatively increased sensitivity to low-density regions.  The degeneracy between density and temperature can be broken by combining other measurements such as X-ray spectral measurements. However, the measurement is challenging due to the relative weakness of the signal and the low resolution of available tSZ maps: The \planck satellite provides a reliable map of tSZ signal with the full-sky coverage and high sensitivity \citep{Planck2016-XXII} but with only moderate resolution (10 arcminute beam).

Luminous red galaxies (LRGs) are powerful tracers of the large-scale structure of the Universe.  These early-type, massive galaxies, selected on the basis of color and magnitude, have mainly old stellar populations with little ongoing star formation.  LRGs typically reside in the centres of galaxy groups and clusters and have been used to detect and characterize the remnants of baryon acoustic oscillations (BAO) at low to intermediate redshift \citep{Eisenstein2005, Kazin2010, Anderson2014}. 

\cite{Planck2013IR-XI} detected the tSZ signal from low-mass haloes as low as $M_{\rm h} \sim 2 \times 10^{13} \msun$ by stacking the \planck tSZ map around locally brightest galaxies (LBGs) constructed from SDSS DR7 galaxies. \cite{Vikram2017} and \cite{Hill2018} cross-correlated the Planck tSZ map with the SDSS DR4 and DR7 group catalogue from \cite{Yang2007} respectively and measured the tSZ signal with high signal-to-noise over a wide range of objects with $M_{\rm h} \sim 10^{11.5 - 15.5} h^{-1} \msun$.

Surprisingly, \cite{Planck2013IR-XI} found that the scaling relation between the integrated tSZ signal and mass follows a simple self-similar relation down to halo masses as low as $M_{\rm h} \sim 2 \times 10^{13} \msun$, apparently indicating that non-gravitational effects are minor even in low-mass haloes.  A consistent result was derived by \cite{Greco2015} using aperture photometry, as opposed to the matched filter technique employed in the \cite{Planck2013IR-XI} study.  These results effectively imply that the gas fraction is approximately independent of halo mass over the large range of halo masses sampled.  However, direct resolved X-ray observations of galaxy groups and clusters (e.g., \citealt{Gastaldello2007, Pratt2009, Sun2009, Gonzalez2013}) have consistently shown that galaxy groups are significantly deficient in their gas content compared to massive clusters.  Using cosmological hydrodynamical simulations that include AGN feedback and which reproduce the properties of local X-ray groups and clusters, \cite{Brun2015} offered a possible solution to this conundrum.  Namely, the relatively coarse resolution of the Planck tSZ map effectively prevents a robust measurement of the tSZ flux on scales of $\la r_{500}$, which is the region the X-ray observations are generally confined to.  \cite{Brun2015} demonstrated that they could recover the inferred self-similar result when the simulations were convolved with the Planck beam and analysed in the same way as the real data.  The upshot of that study is that, when measured within $r_{500}$, the gas properties (particularly the gas fraction) of groups and clusters are not self-similar.  However, the self-similar scaling is recovered on larger scales, which are well sampled by Planck.

The studies mentioned above focused on the integrated tSZ flux within some aperture. However, with the advent of large, publicly-available tSZ maps, it is also important to study how the tSZ signal (and therefore electron pressure) is spatially distributed around galaxies/haloes. For example, \cite{Hill2018} measured the tSZ--galaxy group cross-correlation function and modelled it including signals from correlated halos (`two-halo' term), which was neglected in the \cite{Planck2013IR-XI} study, and found moderate evidence of deviation from self-similarity in the pressure -- mass relation. In this way, comparisons of the spatial distribution to models as well as simulations can provide a potentially strong test of their realism and to deduce the importance of particular processes (e.g., gravitational shock heating vs.~AGN feedback).  The aim of the present study is to do just this.  Specifically, we derive the stacked radial tSZ distribution, $y(\theta)$, around LRGs and we compare it to the predictions of cosmological hydrodynamical simulations and a simple analytic halo model that adopts the so-called `universal pressure profile' \citep{Arnaud2010, Planck2013IR-V} with a significant contribution from nearby clustered haloes. 

Throughout this work, we adopt a $\Lambda$CDM cosmology with parameters from the \cite{Planck2014-I} data release. All masses are quoted in Solar mass and $M_{\Delta}$ is the mass enclosed within a sphere of radius $R_{\Delta}$ such that the enclosed density is $\Delta$ times the \emph{critical} density at redshift $z$. 

This paper is set out as follows: In Section \ref{sec:model} we describe a model to predict the tSZ signal around LRGs. In Section \ref{sec:data}, we summarize the data sets used in our analysis: the SDSS DR7 LRG catalogue, \planck $y$ map and the cosmo-OWLS suite of hydrodynamic simulations. In Section \ref{sec:stacking}, we employ a stacking method to measure the average structure around LRG haloes since the signal-to-noise ratio of the \planck $y$ map is not high enough to trace individual haloes. Our result is compared with the cosmo-OWLS simulations, some of which include AGN feedback, in Section \ref{sec:comp-sim} and we compare to semi-analytical model predictions in Section \ref{sec:comp-model}. In Section \ref{sec:systemaics}, we discuss possible systematic errors in our measurements. Finally, we discuss the interpretation of our findings in Section \ref{sec:discussion} and summarize them in Section \ref{sec:conclusion}.

\section{Basic formalism}
\label{sec:model}

\subsection{The thermal SZ effect}
The tSZ effect is a distortion of the CMB spectrum produced by the inverse Compton scattering of CMB photons off hot electrons along the line of sight, e.g., by ionized gas in the ICM. The change to the CMB temperature, $\Delta T$, at frequency $\nu$ in an angular direction of $\mathbf{\hat{n}}$ is given by 
\beq
    \frac{\Delta T}{T}(\nu,\mathbf{\hat{n}}) = f \left( \frac{h \nu}{k_{\rm B} T} \right) y(\mathbf{\hat{n}})\ , 
\label{eq1}
\eeq
where $k_{\rm B}$ is the Boltzmann constant, $h$ is the Planck constant, and $T$ is the temperature of the CMB. The frequency dependence of the effect is restricted to the pre-factor $f$, where
\beq
    f(x) = x \ \mathrm{coth} \left(\frac{x}{2} \right) -4\ ,
\label{eq2}
\eeq
while the Compton $y$ parameter contains the angular dependence. The Compton $y$ parameter is proportional to the line-of-sight integral of electron pressure, $P_{\rm e}=n_{\rm e}k_{\rm B}T_{\rm e}$. Here $n_{\rm e}$ is the physical electron number density and $T_{\rm e}$ is the electron temperature. The line-of-sight integral is:
\beq
    y(\mathbf{\hat{n}}) = \frac{\sigma_{\rm T} }{m_{\rm e} {\it c}^2}
	\int P_{\rm e}(l,\mathbf{\hat{n}}) \, \der l\ ,
\label{eq3}
\eeq
where $\sigma_{\rm T}$ is the Thomson cross section, $m_{\rm e}$ is the mass of electron, $c$ is the speed of light and $l$ is the \emph{physical} distance. We ignore relativistic corrections to the tSZ spectrum (\eg \citealt{Itoh1998}), which only become non-negligible for the most massive clusters of $\gtrsim 10^{15} \msun$. 

\subsection{The stacked $y$ profile}
\label{sec:gy-cross}
For the calculation of the stacked $y$ profile, we follow the method in \cite{Fang2012} and work in the flat-sky and Limber approximation \citep{Limber1954}. 

The cross power spectrum for the tSZ signal and the distribution of galaxy clusters is given by the sum of a `one-halo term', which counts correlation arising within an individual halo, and a `two-halo term', which counts correlation arising due to the environment surrounding a halo \citep{Komatsu2002, Cooray2002}:
\beq
    C^{yh}_{\ell} = C^{yh,{\rm 1h}}_{\ell} + C^{yh, {\rm 2h}}_{\ell}\ . 
\label{eq4}
\eeq
The one-halo term is given by
\beq
\begin{aligned}
    C^{yh,{\rm 1h}}_{\ell} = \frac{1}{\bar{n}^{\rm 2D}} \int \der z \frac{\der^2 V}{\der z \der \Omega} \int \der M \frac{\der n}{\der M}(M,z) \\ 
	\times S(M,z) \tilde{y}_{\ell}(M,z)\ , 
\end{aligned}
\label{eq5}
\eeq
where $\der^2 V / \der z \der \Omega$ is the comoving volume element per redshift per steradian and $\der n/\der M$ is the halo mass function (sometimes denoted $n(M,z)$ in the literature; the comoving number density of haloes in a mass interval $\der M$). We adopt the halo mass function of \cite{Tinker2010} and use `HMFcalc\footnote{\href{http://hmf.icrar.org/}{http://hmf.icrar.org/}}' \citep{Murray2013} for the calculation. The selection function, $S(M,z)$, defines the redshift and halo mass. In our work, the halo masses of LRGs are estimated using stellar-to-halo masss (SHM) relations in Section \ref{subsec:halomass}, which are applied to the stellar mass distribution of LRGs in Fig.~\ref{f01}. The average two-dimensional angular number density of the selected haloes is calculated via
\beq
    \bar{n}^{\rm 2D} = \int \der z \frac{\der^2 V}{\der z \der \Omega} \int \der M \frac{\der n}{\der M}(M,z) S(M,z)\ . 
\label{eq7}
\eeq

Here $\tilde{y}_{\ell}(M,z)$ is the 2D Fourier transform of the $y$ profile for a halo with a pressure profile $P_{\rm e}(x,M,z)$, given by 
\beq
    \tilde{y}_{\ell}(M,z) = \frac{\sigma_{\rm T} }{m_{\rm e} {\it c}^2} \frac{4 \pi r_{\rm s}}{\ell^{2}_{\rm s}} \int \der x x^{2} \frac{\sin (\ell x/\ell_{\rm s})}{\ell x/\ell_{\rm s}} P_{\rm e}(x,M,z)\ , 
\label{eq8}
\eeq
where 
\beq
    x = \frac{r}{r_{\rm s}}, \quad \ell_{\rm s} = \frac{d_{\rm A}}{r_{\rm s}}\ , 
\label{eq9}
\eeq
and where $r_{\rm s}$ is the characteristic scale radius of the pressure profile, $x = r/r_{\rm s}$ is the dimensionless radial scale, and $d_{\rm A}$ is the angular diameter distance. $\ell_{\rm s} = d_{\rm A}/r_{\rm s}$ is the associated multipole moment. 
The two-halo term is given by
\beq
    C^{yh,{\rm 2h}}_{\ell} = \int \der z \frac{\der^2 V}{\der z \der \Omega}  W^{\rm h}(z) W^{y}_{\ell}(z)P^{\rm L}_{\rm m}\left(k=\frac{\ell}{\chi},z \right)\ , 
\label{eq10}
\eeq
where $P^{\rm L}_{\rm m}(k,z)$ is the linear matter power spectrum. The function $W^{\rm h}(z)$ is defined as 
\beq
    W^{h}(z) = \frac{1}{\bar{n}^{\rm 2D}} \int \der M \frac{\der n}{\der M}(M,z) S(M,z) b(M,z)\ ,
\label{eq11} 
\eeq
and $W^{y}_{\ell}(z)$ is 
\beq
    W^{y}_{\ell}(z) = \int \der M \frac{\der n}{\der M}(M,z) b(M,z) \tilde{y}_{\ell}(M,z)\ , 
\label{eq12}
\eeq
where $b(M,z)$ is the halo bias. We take the halo bias from \cite{Tinker2010}. 

By summing the two- and one-halo terms together, the Fourier-transform of the stacked $y$ profile, $C^{yh}_{\ell}$, can be calculated. In our work we are interested in comparing our model to the angular configuration space stacked $y$ profile, which can be obtained from our model via an inverse Fourier transform:
\beq
    \bar{y} (\theta) = \int \frac{\ell \der\ell}{2\pi} J_{0}(\ell \theta) C^{yh}_{\ell}\ , 
\label{eq13}
\eeq
where $J_{0}$ is the zeroth order Bessel function. Finally, we convolve our model with the point-spread function of the \planck beam 
\beq
    \bar{y} (\theta)_{\rm obs} = \int \frac{\ell \der\ell}{2\pi} J_{0}(\ell \theta) C^{yh}_{\ell} B_{\ell}\ , 
\label{eq14}
\eeq
where $B_{\ell}=\exp[-\ell(\ell+1)\sigma^2/2$] and $\sigma = \theta_{\rm FWHM}/\sqrt{8 \, \mathrm{ln}(2)}$ with $\theta_{\rm FWHM} = 10$ arcmin, which corresponds to the beam of the \planck $y$ map.

\subsection{The universal pressure profile}
For the electron pressure profile, we adopt the `universal' pressure profile \citep[UPP;][]{Nagai2007}, which is a form of generalized \citeauthor*{Navarro1997} (NFW; \citeyear{Navarro1997}) profile, 
\beq
    \mathbb{P} (x) = \frac{P_{0}}
	{(c_{500} x)^{\gamma} [1 + (c_{500} x)^{\alpha}]^{(\beta-\alpha)/\gamma} }\ .
\label{eq15}
\eeq
Here, $x = r/R_{500}$ and we remind the reader that $R_{500}$ relates to $500$ times the critical density. The model is defined by the following parameters: $P_{0}$, normalization; $c_{500}$, concentration parameter defined at a characteristic radius $R_{500}$; and the slopes in the central $(x \ll 1/c_{500})$, intermediate $(x \sim 1/c_{500})$ and outer regions $(x \gg 1/c_{500})$, given by $\gamma$, $\alpha$ and $\beta$, respectively. The scaled pressure profile for a halo with $M_{500}$ and $z$ is
\beq
    \frac{P(r)}{P_{500}} = \mathbb{P}(x), 
\label{eq16}
\eeq
with 
\beq
\begin{aligned}
    P_{500} & = 1.65 \times 10^{-3} \left[\frac{H(z)}{H_0}\right]^{8/3}	\\
	& \times \left[ \frac{ (1-b) \, M_{500}}{3 \times 10^{14} \, (h/0.7)^{-1} \msun} \right]^{2/3} 
	\left(\frac{h}{0.7}\right)^{2} \rm \ keV \ cm^{-3}, 
\end{aligned}
\label{eq17}
\eeq
where $H(z)$ is the Hubble parameter at redshift $z$ and $H_0=100h\,\mathrm{km s^{-1} Mpc^{-1}}$ is the present value.  $P_{500}$ is the characteristic pressure reflecting the mass variation expected in a self-similar model of pressure evolution, purely based on gravitation \citep{Arnaud2010}.  Note that $M_{500}$ is the `true' mass from lensing measurements in this paper and $(1-b)$ is the hydrostatic mass bias, and this hydrostatic mass, $(1-b) \, M_{500}$, corresponds to $M_{500}$ in \cite{Arnaud2010}. For the mass bias, we adopt $ (1-b) \simeq 0.78$ derived from the Canadian Cluster Comparison Project \citep{Hoekstra2015}. Deviation from self-similar scaling appears as a variation of the scaled pressure profile and, as in \cite{Arnaud2010}, this variation is expressed as a function of $M_{500}$, 
\beq
    \frac{P(r)}{P_{500}} = 
	\mathbb{P}(x) \left[ \frac{ (1-b) \, M_{500}}{3 \times 10^{14} \, (h/0.7)^{-1} \msun} \right]^{\alpha_{\rm p}}, 
\label{eq18}
\eeq
where $\alpha_{\rm p} = 0.12$. For the parameters of the generalised NFW electron pressure profile, we adopt the best-fit values of $[P_{0},c_{500}, \gamma, \alpha, \beta] = [6.41,1.81,0.31,1.33,4.13]$, estimated from 62 massive nearby clusters ($10^{14.4} < M_{500} < 10^{15.3} \msun$) using the \planck tSZ and \xmm X-ray data in \cite{Planck2013IR-V}. The deviation from the self-similar relation ($\alpha_{\rm p}$) is likely driven by the fact that the gas mass fraction varies with halo mass, with low-mass haloes having lower gas fractions.  X-ray observations suggests higher value of $\sim 0.26$ for the deviation using galaxy groups/clusters with $10^{13} < M_{500} < 10^{15} \msun$ in \cite{Gonzalez2013}. A consistent result is obtained in \cite{Anderson2015} using the scaling relation, $L_{\rm X}$-$M_{500}$, of haloes with $10^{12.6} < M_{500} < 10^{14.6} \msun$. We will test it with the spatial distribution of pressure ($y$) including a contribution from nearby clustered haloes.

%
%

\section{Data}
\label{sec:data}

We use three data sets in this analysis: the Luminous Red Galaxy catalogue from the Sloan Digital Sky Survey seventh data release\footnote{http://cosmo.nyu.edu/~eak306/SDSS-LRG.html} (SDSS DR7 LRG, N$_{\rm LRG}$=105,831, \citealt{Kazin2010}), the \planck Comptonization $y$ map\footnote{http://pla.esac.esa.int/pla/\#results} from the 2015 data release \citep{Planck2016-XXII} and the cosmo-OWLS suite of cosmological hydrodynamic simulations \citep{Brun2014, Daalen2014, McCarthy2014}.  We describe each briefly in the following subsections.

\subsection{Luminous Red Galaxy catalogue}
\label{lrg}

\begin{figure*}
\begin{center}
\begin{minipage}{0.49\linewidth}
\includegraphics[width=\linewidth]{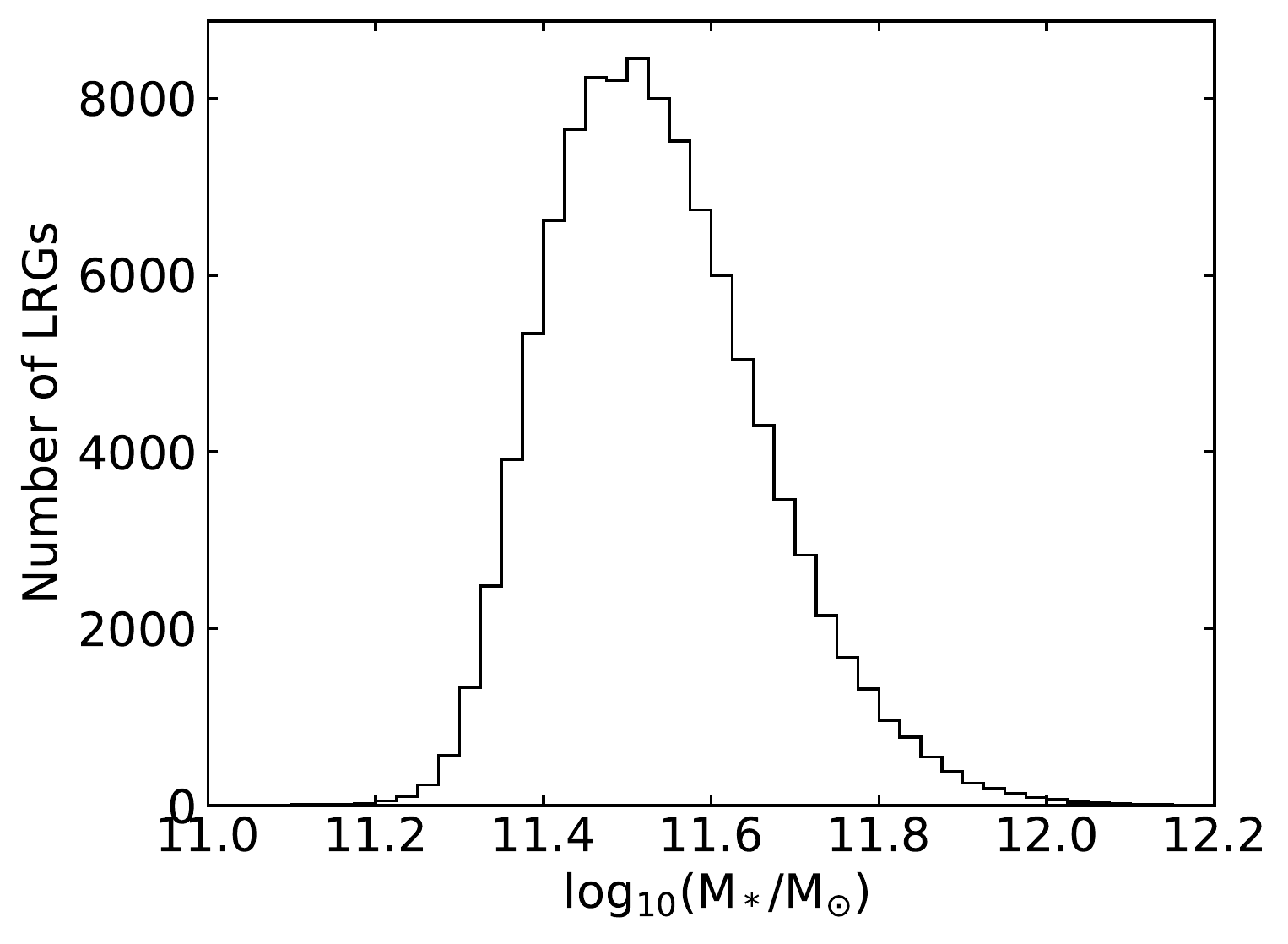}
\end{minipage}
\begin{minipage}{0.49\linewidth}
\includegraphics[width=\linewidth]{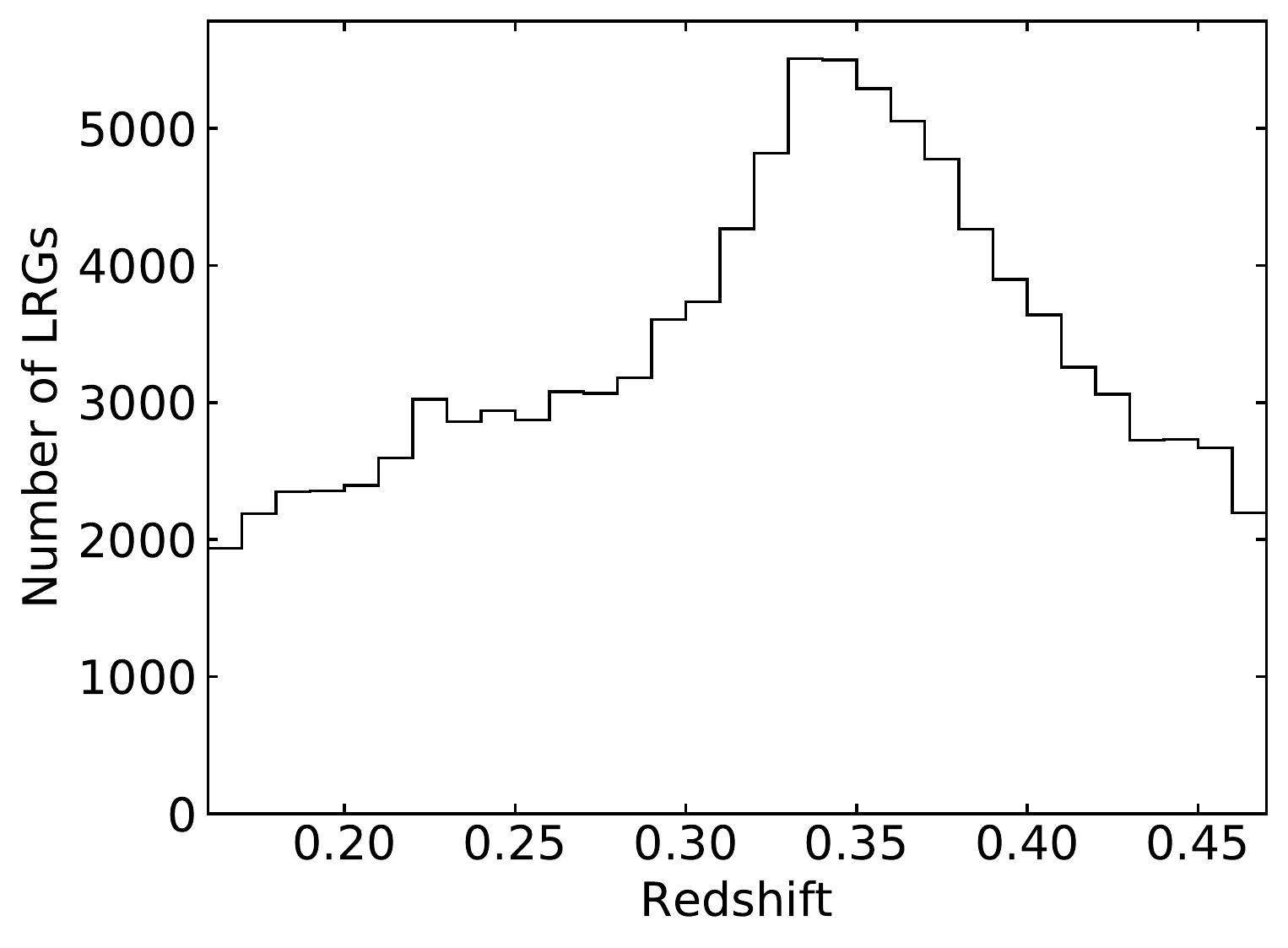}
\end{minipage}
\end{center}
\caption{{\it Left}: The stellar mass distribution of SDSS DR7 LRGs. 
{\it Right}: The redshift distribution of the LRGs. }
\label{f01}
\end{figure*}

The LRG catalogue provides galaxy positions, magnitudes and spectroscopic redshifts. Stellar masses of the LRGs are provided in the New York University Value-Added catalogue (NYU-VAGC)\footnote{http://sdss.physics.nyu.edu/vagc/} \citep{Blanton2005}, which are estimated with the K-correct software\footnote{http://howdy.physics.nyu.edu/index.php/Kcorrect} of \cite{Blanton2007} by fitting the five-band SDSS photometry to more than 400 spectral templates. Most of the templates are based on stellar evolution synthesis models of \cite{Bruzual2003} assuming the stellar initial mass function of \cite{Chabrier2003}. The stellar masses in the NYU-VAGC catalogue are given in a unit of $\msun h^{-2}$ and we take $h=0.671$ from the \planck cosmology (Fig.~\ref{f01}).

\begin{table*}
\centering
\caption{The baryon feedback models in the cosmo-OWLS simulations. Each model has been run with both \planck and \wmap cosmological parameters.}
\begin{tabular}{|l|l|l|l|l|l|l|} \hline
Simulation & UV/X-ray background & Cooling & Star formation & SN feedback & AGN feedback & $\Delta T_{\rm heat}$ \\ \hline
NOCOOL & Yes & No & No & No & No & ... \\ 
REF & Yes & Yes & Yes & Yes & No & ... \\ 
AGN 8.0 & Yes & Yes & Yes & Yes & Yes & $10^{8.0}$ K \\ 
AGN 8.5 & Yes & Yes & Yes & Yes & Yes & $10^{8.5}$ K \\
AGN 8.7 & Yes & Yes & Yes & Yes & Yes & $10^{8.7}$ K \\ \hline
\end{tabular}
\label{tab1}
\end{table*}

\subsection{Planck $y$ map}
The \planck tSZ map is one of the datasets provided in the \planck 2015 data release.  The map comes in HEALPix\footnote{http://healpix.sourceforge.net/} \citep{Gorski2005} format   with a pixel resolution of $N_{\rm side}$ = 2048.  Two types of $y$ maps are publicly available: MILCA \citep{hurier2013} and NILC \citep{remazeilles2013}, both of which are based on multi-band combinations of the \planck band maps \citep{Planck2016-I}.  Our analysis is based on the MILCA map, but we obtain consistent results if we use the NILC map.

The 2015 data release also provides sky masks suitable for analyzing the $y$ maps, including a point-source mask and galactic masks of varying severity: masking 40, 50, 60 or 70\%  of the sky.  We combine the point source mask with the 40\% galactic mask, which excludes $\sim$50\% of the sky.   The mask is applied during the stacking process: for a given LRG, masked pixels in the $y$ map near that LRG are not accumulated in the stacked image. We accept the 77,762 LRGs for which 80\% of the region within a 40 arcmin circle around each LRG is available. We reject the others in case the mask may bias the measured $y$ profile. 

\subsection{Simulations}
\label{cosmo-owls}

To compare our results with theory, we analyze the cosmo-OWLS suite of cosmological smoothed particle hydrodynamics (SPH) simulations \citep{Brun2014, Daalen2014, McCarthy2014} in the same manner as the data.  The cosmo-OWLS suite is an extension of the OverWhelmingly Large Simulations project \citep{Schaye2010} designed with cluster cosmology and large-scale structure surveys in mind (see also \citealt{McCarthy2017}). The cosmo-OWLS suite consists of box-periodic hydrodynamical simulations,  the largest of which have volumes of (400$h^{-1} \mathrm{Mpc}$)$^3$ and contain $1024^3$ each of baryonic and dark matter particles. The suite employs two different cosmological models: the \planck 2013 cosmology \citep{Planck2014-I} with
\begin{align*}
\qquad & \lbrace \Omega_{\rm m}, \Omega_{\rm b}, \Omega_{\Lambda}, \sigma_8 , n_{\rm s} , h \rbrace = \\
& \lbrace 0.3175, 0.0490, 0.6825, 0.834, 0.9624, 0.6711 \rbrace, 
\end{align*} 
and 
the \wmap cosmology \citep{Komatsu2011} with
\begin{align*}
\qquad & \lbrace \Omega_{\rm m}, \Omega_{\rm b}, \Omega_{\Lambda}, \sigma_8 , n_{\rm s} , h \rbrace = \\ 
& \lbrace 0.272, 0.0455, 0.728, 0.81, 0.967, 0.704 \rbrace. 
\end{align*}
Each simulation is run with 5 different models of baryon sub-grid physics: `NOCOOL', `REF', `AGN 8.0', `AGN 8.5' and `AGN 8.7', which are summarized in Table.~\ref{tab1}.

NOCOOL is a standard non-radiative adiabatic model that includes hydrodynamical baryons, but does not produce stars. REF is the OWLS reference model including UV/X-ray background, radiative cooling, star formation and stellar feedback. The AGN models are built on the REF model, and that additionally includes black hole growth and feedback from active galactic nuclei. The three AGN models differ only in their choice of the key parameter of the AGN feedback model $\Delta T_{\rm heat}$, which is the temperature by which neighbouring gas is raised due to feedback. Increasing the value of $\Delta T_{\rm heat}$ results in more energetic feedback events, and also leads to more bursty feedback, since the black holes must accrete more matter in order to heat neighbouring gas to a higher adiabat. Earlier studies demonstrate that the AGN 8.0 model reproduces a variety of observed gas features in local groups and clusters of galaxies by optical and X-ray data (e.g., \citealt{Brun2014}). 

For each simulation, 10 quasi-independent mock galaxy catalogues are generated on 10 light cones and 10 corresponding $y$ maps are generated from periodic boxes of randomly rotated and translated simulation snapshots (redshift slices) along the line-of-sight back to $z$ = 3 \citep{McCarthy2014}. Each of these light cones contain about one million galaxies and each spans a $5^{\circ} \times 5^{\circ}$ patch of sky.  To compare with data, we convolve the simulated $y$ maps with a Gaussian kernel of 10 arcmin in FWHM, corresponding to the beam of the \planck $y$ map.

\section{Stacking $y$ map centred on LRG\lowercase{s}}
\label{sec:stacking}

In this section, we describe our procedure for stacking the \planck $y$ map against the LRGs and for constructing the mean $y$ profile: We place each LRG in our catalogue at the centre of a 2-dimensional angular coordinate system of $-40' < \Delta l < 40'$ and $-40' < \Delta b < 40'$ divided into 80 $\times$ 80 bins. We then linearly interpolate the $y$ map onto our grid. For each LRG we subtract the mean tSZ signal in the annular region between 30 and 40 arcmin as an estimate of the local background signal for that particular LRG. Finally we stack all LRGs and then divide by the total number of LRGs in our sample.

We assess the uncertainties in our measurements through bootstrap resampling.  We draw a random sampling of LRGs with replacement and re-calculate an average $y$ value for the new set of LRGs. We repeat this process 1000 times and the bootstrapped data produce 1000 average $y$ values. The uncertainties are estimated by their RMS fluctuation.

\begin{figure*}
\begin{center}
\includegraphics[width=0.6\linewidth]{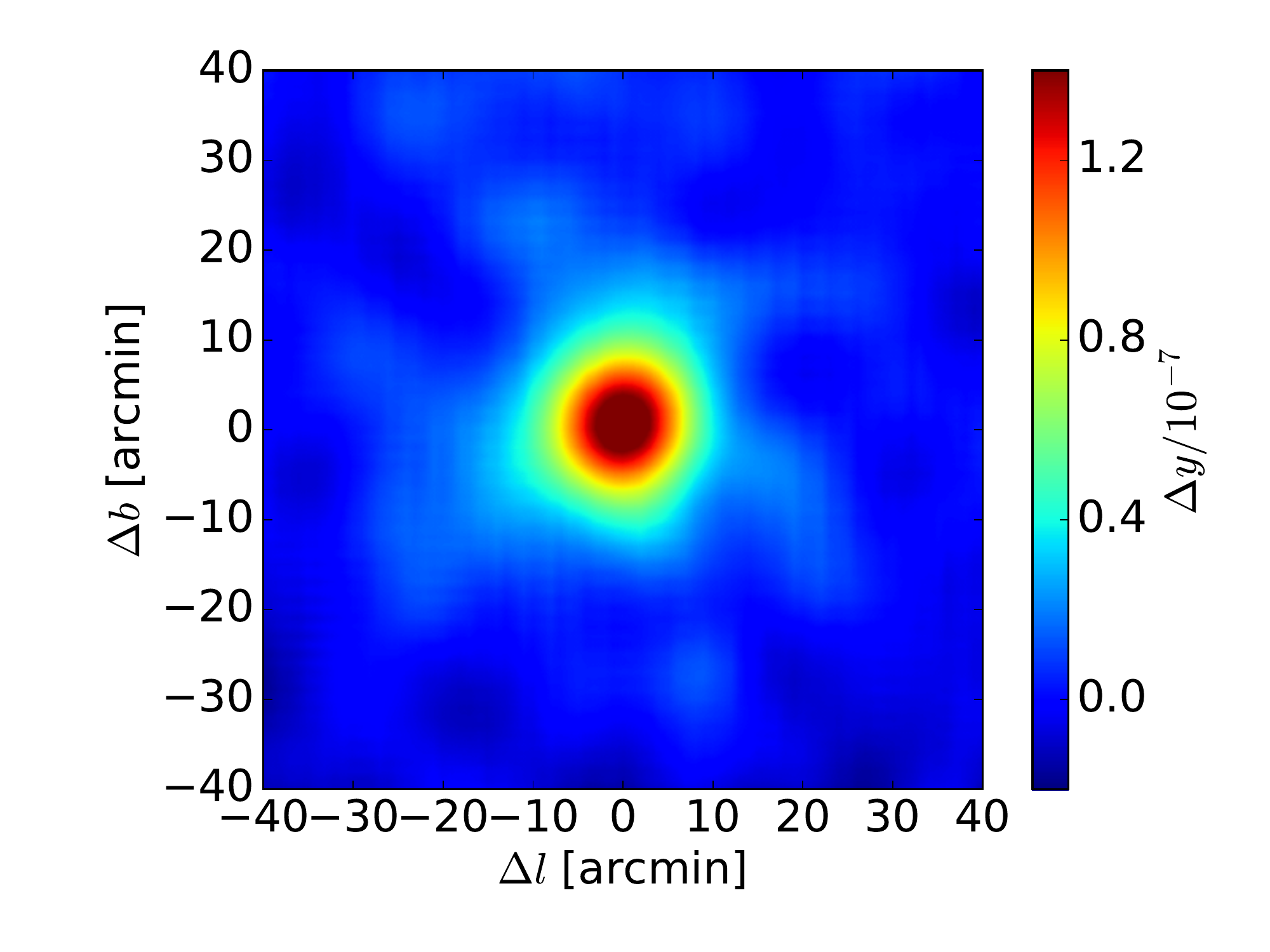}
\includegraphics[width=0.45\linewidth]{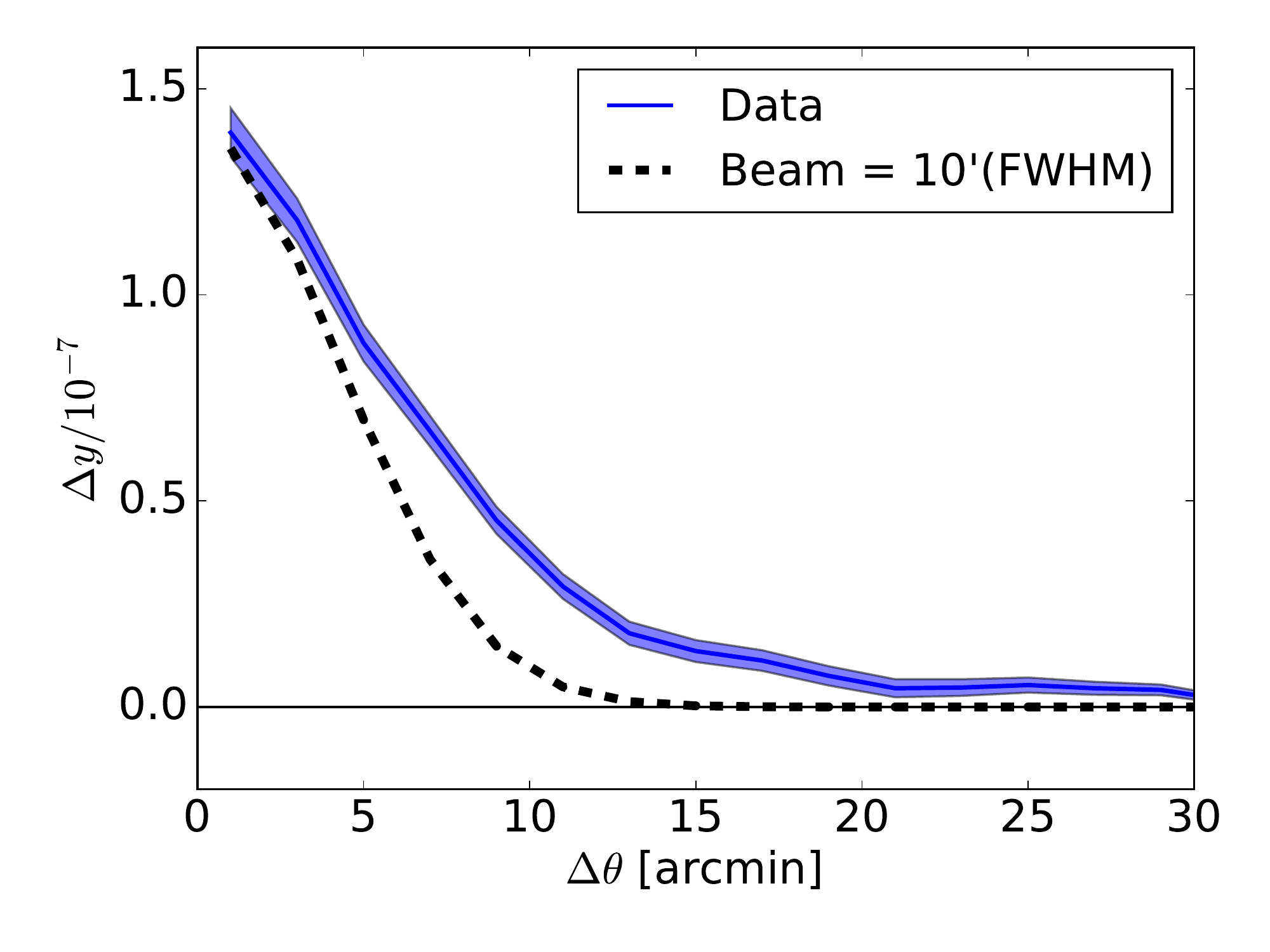}
\includegraphics[width=0.45\linewidth]{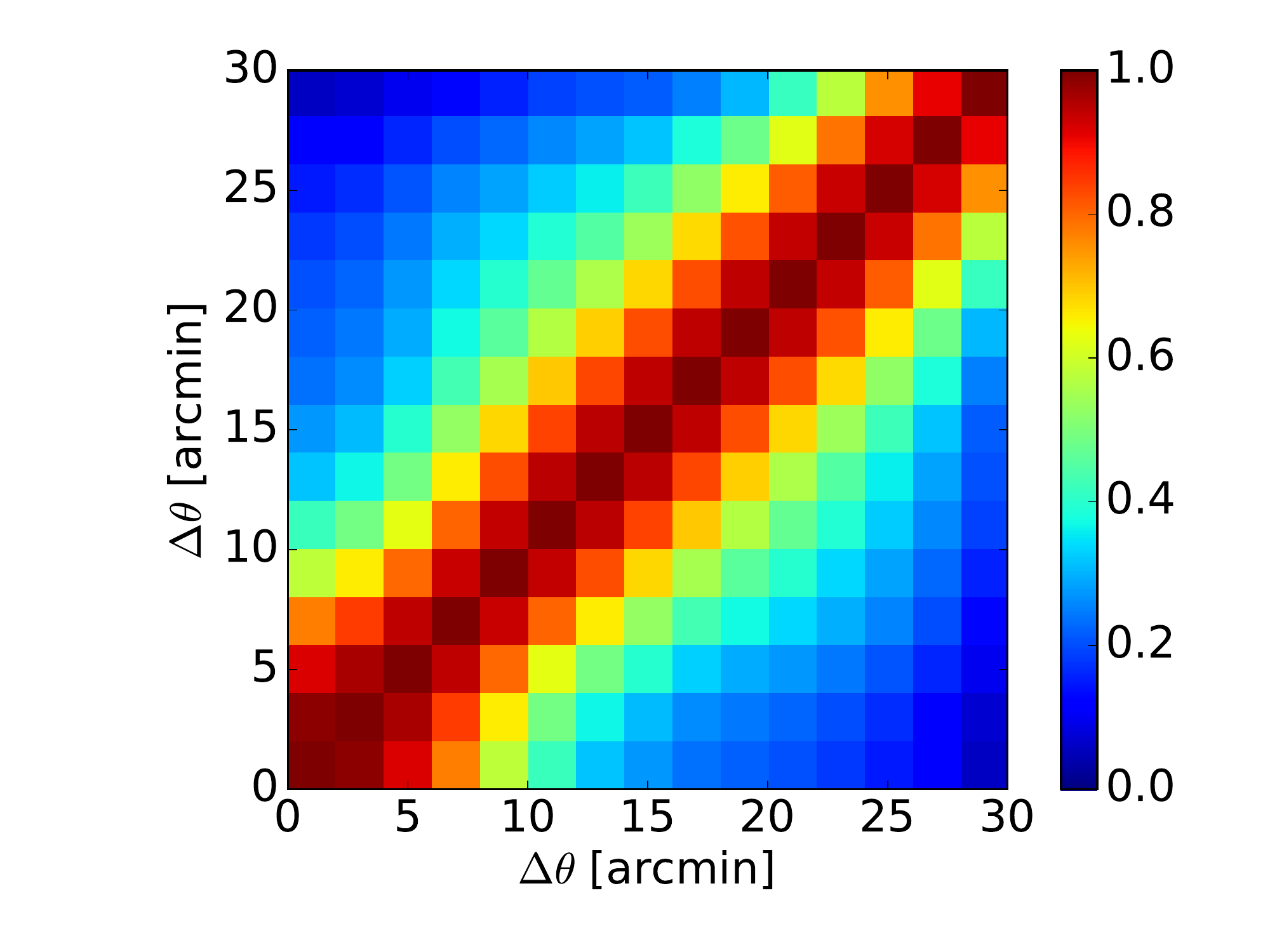}
\end{center}
\caption{
{\it Top}: The average \planck $y$ map stack, centred on 77,762 LRGs in an angular coordinate system of $-40' < \Delta l < 40'$ and $-40' < \Delta b < 40'$ divided in 80 $\times$ 80 bins. {\it Bottom left}: The average $y$ profile around the LRGs. The 1$\sigma$ statistical uncertainty is represented via the width of the blue line. The $\mathrm{FWHM}=10$ arcmin Gaussian beam of of the \planck $y$ map is shown as the black dashed line for comparison, the peak of which is normalized to the centre of the LRGs' $y$ profile. {\it Bottom right}: The correlation matrix between different radial bins of the profile.} 
\label{f02}
\end{figure*}

The top panel in Fig.~\ref{f02} shows the average $y$ map stacked against the 77,762 LRGs. The bottom-left panel in Fig.~\ref{f02} is the average $y$ profile of the LRGs, where width of the blue line represents a 1$\sigma$ statistical uncertainty of the $y$ profile. The 10 arcmin Gaussian beam, normalized to the central peak of the measured $y$ profile, is shown as a black dashed line for comparison. We detect the tSZ signal out to $\sim$ 30 arcmin, well beyond the 10 arcmin beam of the \planck $y$ map. The bottom-right panel of Fig.~\ref{f02} shows the correlation matrix between different radial bins of the profile, in which the effect of the beam is seen. 

\section{Comparison with hydrodynamic simulations}
\label{sec:comp-sim}

\subsection{Estimating halo masses of LRGs} 
\label{subsec:halomass}

\begin{figure}
\begin{center}
\includegraphics[width=\linewidth]{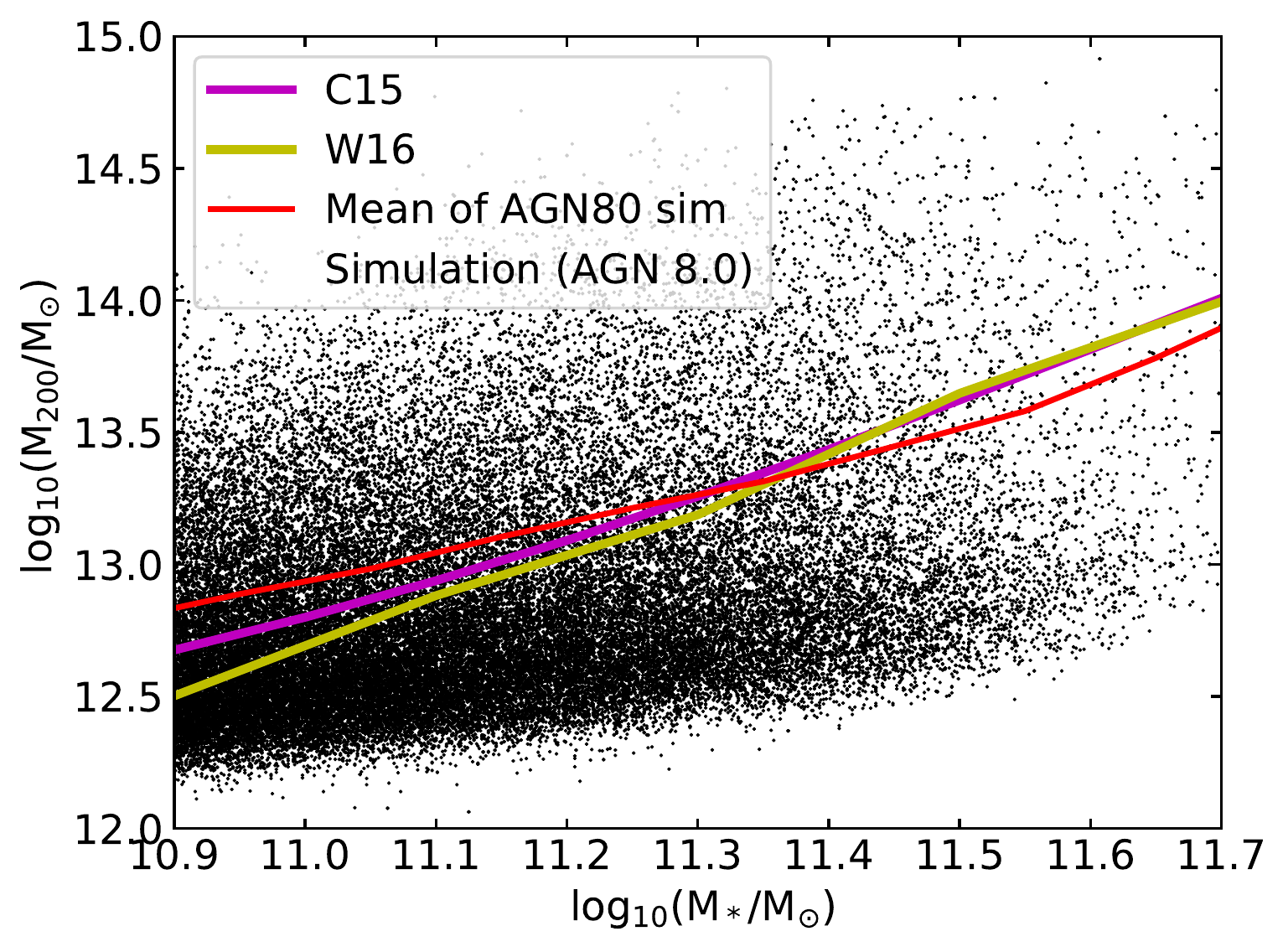}
\caption{Black points show the relation between halo and stellar mass of individual central galaxies with $0.16 < z < 0.47$ from the AGN 8.0 simulation, the mean relation of which is shown in red. There is a large scatter. The three stellar-to-halo mass (SHM) relations for the mean are shown for comparison: C15-SHM \citep{Coupon2015} in magenta and W16-SHM \citep{Wang2016} in yellow. }
\label{f04}
\end{center}
\end{figure}

In order to compare the $y$ profile around the LRGs with simulations, we estimate the halo masses of the LRG haloes using their stellar mass estimates. We do this using the SHM relations from \cite{Coupon2015} (C15-SHM) and \cite{Wang2016} (W16-SHM). In C15-SHM, the relation is estimated in the CFHTLenS/VIPERS field by combining deep observations from the near-UV to the near-IR, supplemented by $\sim$ 70 000 secure spectroscopic redshifts, and analyzing galaxy clustering, galaxy-galaxy lensing and the stellar mass function. In W16-SHM, the SHM relation is estimated for LBGs in \cite{Planck2013IR-XI} by gravitational lensing measurements with a source galaxy catalogue in \cite{Reyes2012}. These empirically-derived SHM relations (C15-SHM in magenta and W16-SHM in yellow) are shown in Fig.~\ref{f04}, along with individual, simulated central galaxies from the cosmo-OWLS AGN 8.0 simulation.  In the stellar mass range of our LRGs, the mean halo mass estimates from C15-SHM and W16-SHM are consistent with each other.  In spite of the large scatter, the AGN 8.0 simulation yields a mean SHM relation (red line in Fig.~\ref{f04}) that is similar to the observed relations.

\subsection{Comparison with simulations} 
\label{subsec:comp-sim}

We now compare the average $y$ profile to that predicted by the simulations.  To do so, we analyse 10 light cones from each hydrodynamic model of the cosmo-OWLS suite of simulations (Section \ref{cosmo-owls}) in exactly the same way as we analyse the real data. To identify simulated LRGs, we select simulated haloes with the same halo mass, estimated in Section \ref{subsec:halomass}, and redshift ranges as in the real data. The average stacked $y$ profile in each mass and redshift bin is then constructed from the simulated light cones. The stacks are then combined, weighted by the total number of LRGs: 
\beq
    y (\theta)_{\rm sim} = \sum_{M_{500},z} \left[  \bar{y} (\theta,M_{500},z)_{\rm sim} \times w(M_{500},z)_{\rm LRG} \right], 
\label{eq19}
\eeq
where $\bar{y} (\theta,M_{500},z)_{\rm sim}$ is the average $y$ profile of simulated haloes in a halo mass, $M_{500}$, and redshift bin, and $w(M_{500},z)_{\rm LRG}$ is the normalized number of actual LRGs in the same halo mass and redshift bin. Since the field-of-view of each light cone (25 deg$^2$) is much smaller than the overlapping region of the SDSS and \planck surveys ($\sim$ 8000 deg$^2$), massive haloes are scarce in the simulations. Due to this scarcity, we restrict the maximum stellar mass of the LRGs (corresponding halo mass) that we take from SDSS in our analysis to $10^{11.7} \msun$ ($M_{500} \sim 10^{14.0} \msun$) in order that we have enough simulated haloes in the mass range of SDSS galaxies. As a result of removing high-mass LRGs, the total number of LRGs available to us is reduced to 66,479. This procedure limits us to LRGs with the stellar mass of $10^{11.2} \leq M_*/\msun \leq 10^{11.7}$, which roughly corresponds to halo masses $10^{13}\leq M_{500}/\msun \leq10^{14}$ as shown in Fig.~\ref{f04}. This is not a great loss, considering that we aim to probe baryonic effects that may be more evident in low-mass group and clusters.

\begin{figure*}
\begin{center}
\begin{minipage}{0.51\linewidth}
\includegraphics[width=\linewidth]{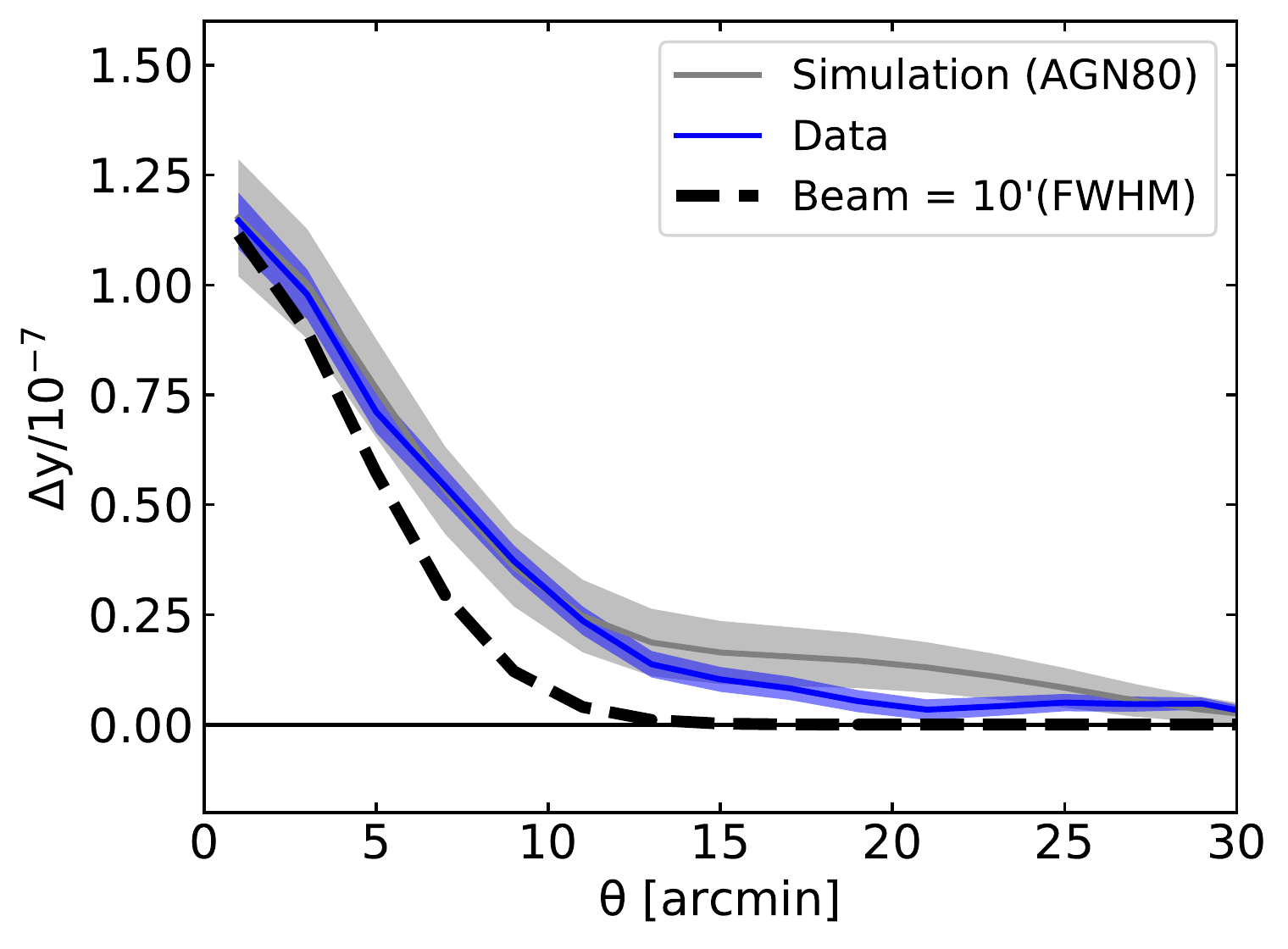}
\end{minipage}
\begin{minipage}{0.49\linewidth}
\includegraphics[width=\linewidth]{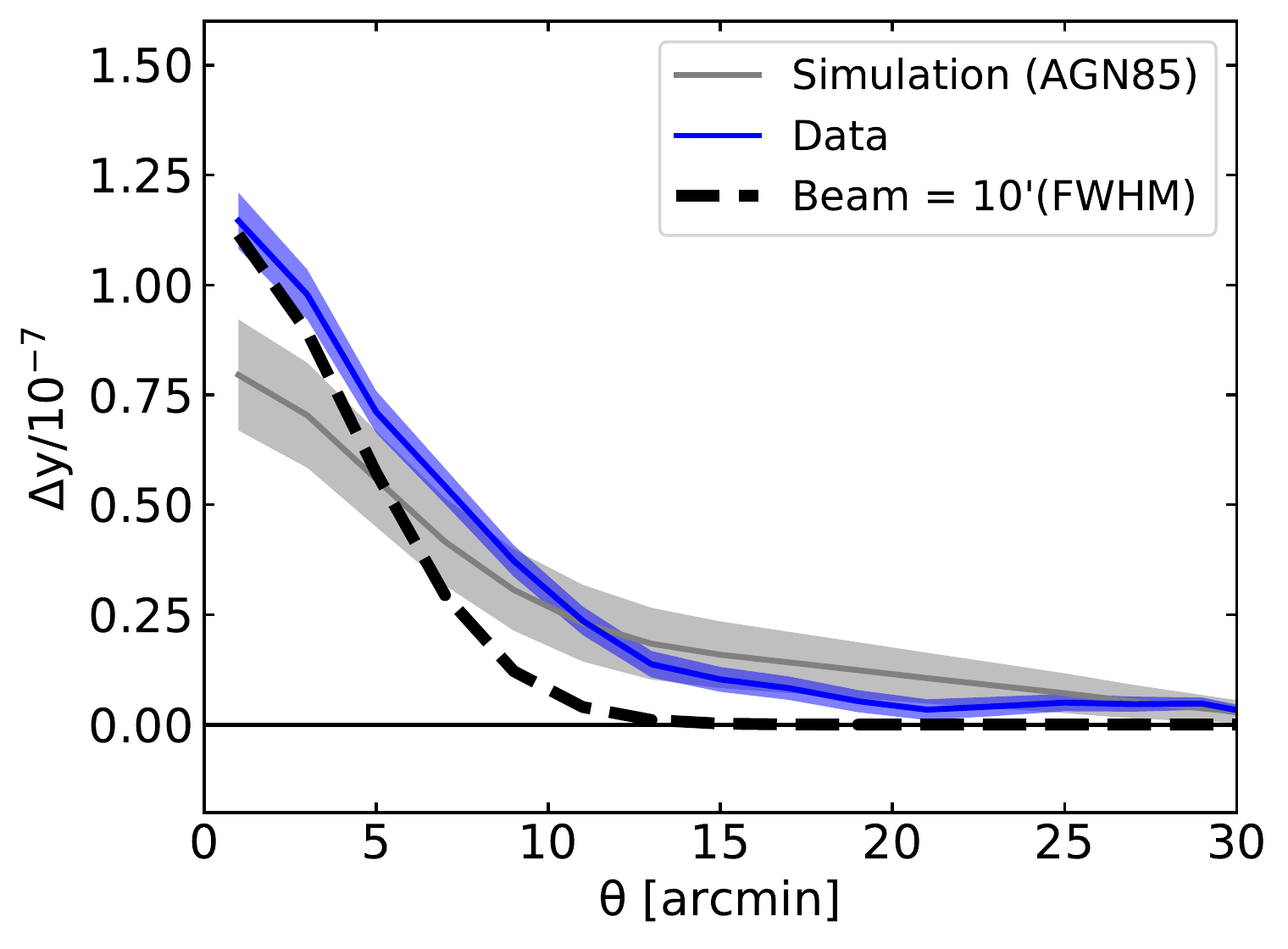}
\end{minipage}
\begin{minipage}{0.49\linewidth}
\includegraphics[width=\linewidth]{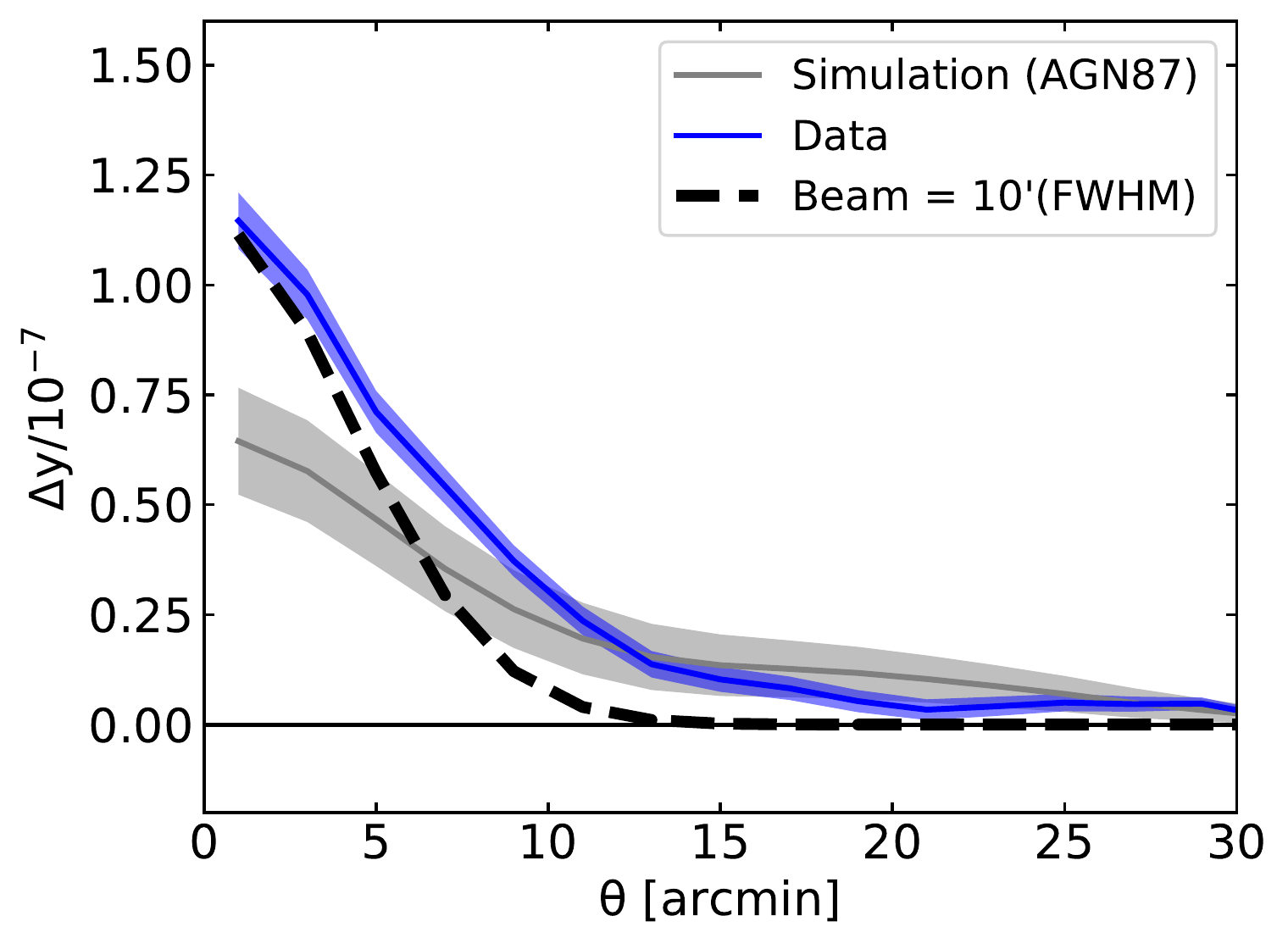}
\end{minipage}
\begin{minipage}{0.49\linewidth}
\includegraphics[width=\linewidth]{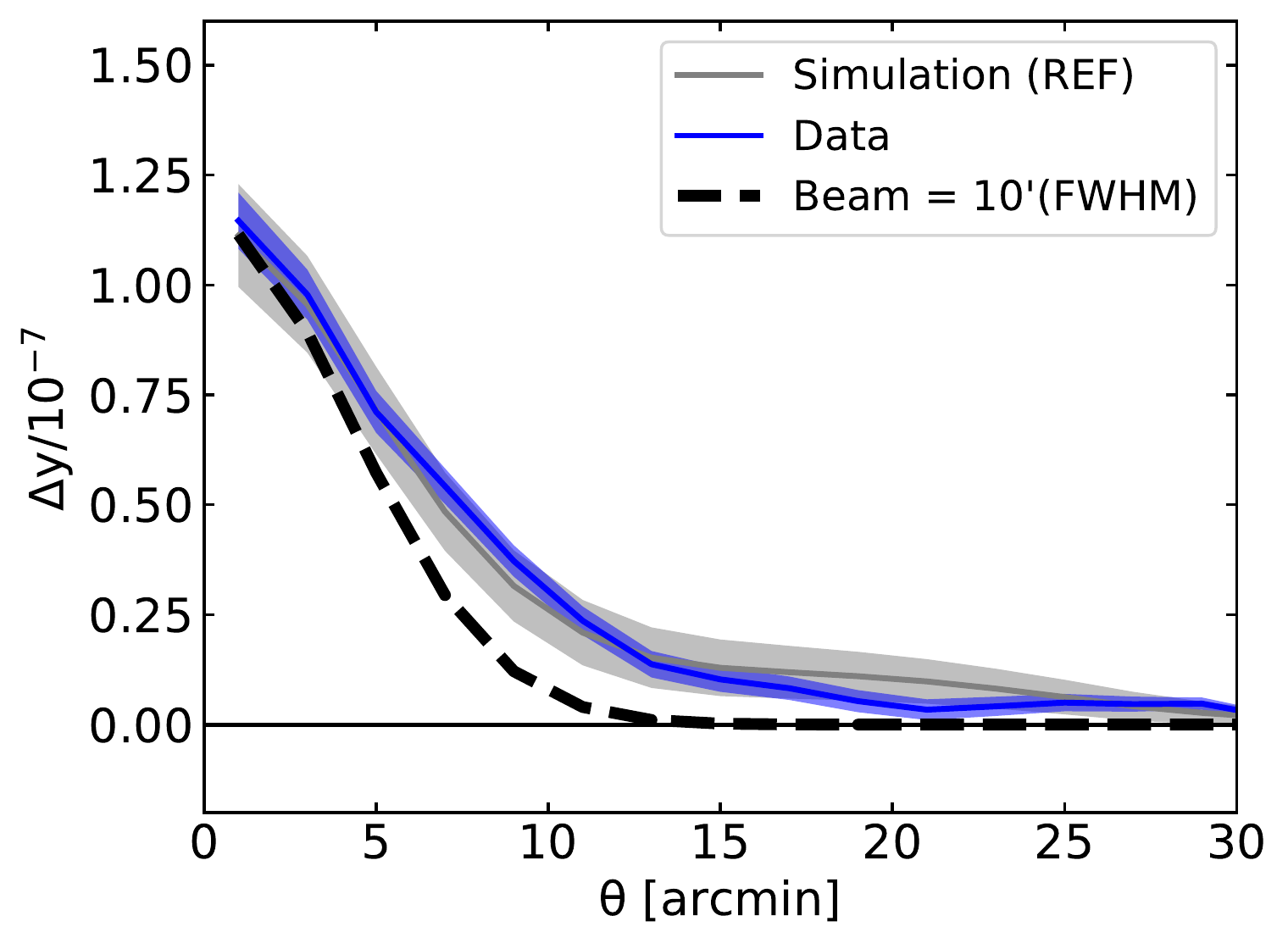}
\end{minipage}
\begin{minipage}{0.49\linewidth}
\includegraphics[width=\linewidth]{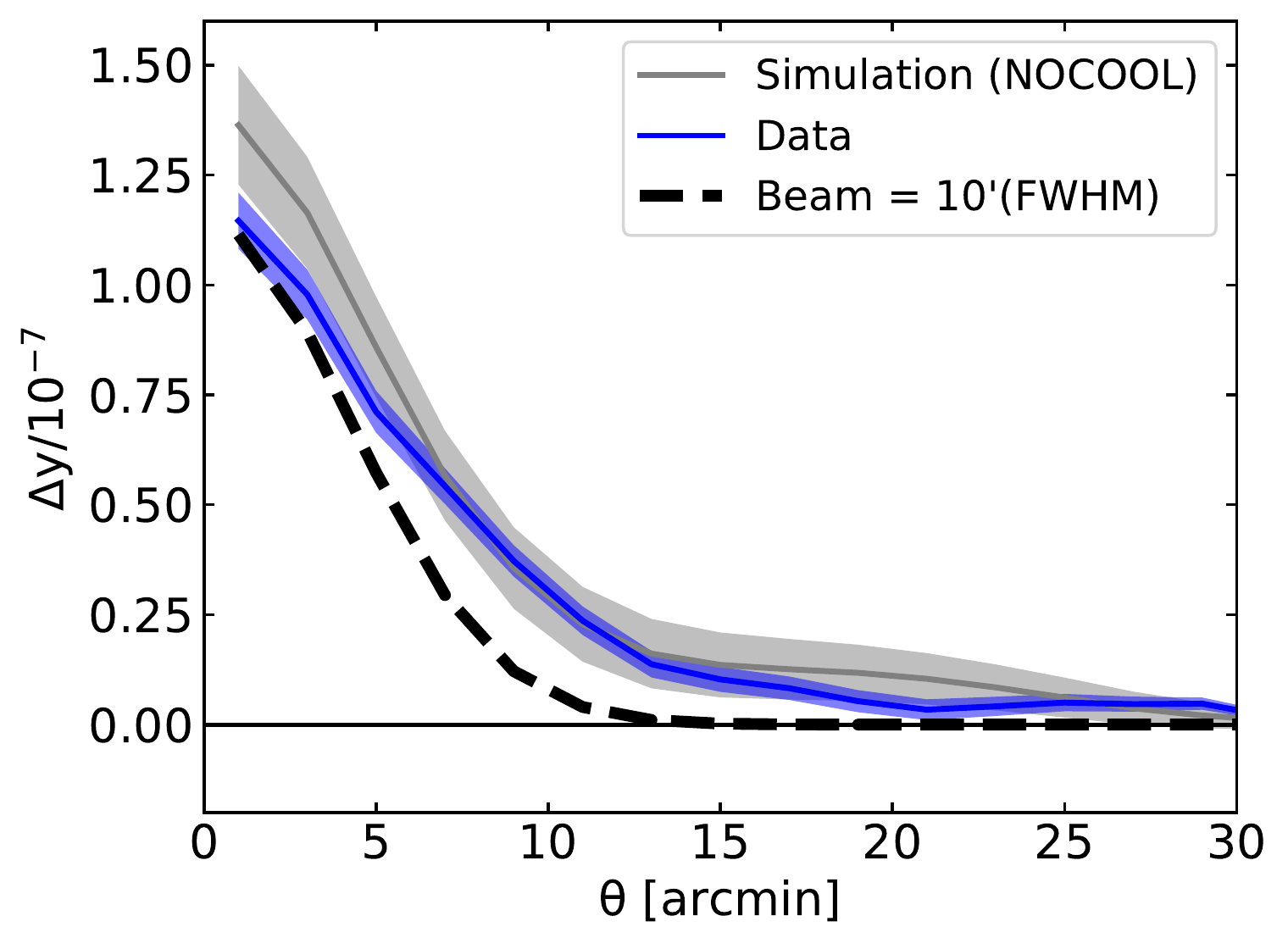}
\end{minipage}
\end{center}
\caption{The average $y$ profile around 66,479 LRGs (blue) is compared to the $y$ profiles of the simulated haloes (gray) in different AGN feedback models respectively. {\it Top}: AGN 8.0 model, {\it Middle left}: AGN 8.5 model, {\it Middle right}: AGN 8.7 model, {\it Bottom left}: REF model and {\it Bottom right}: NOCOOL model. For the comparison, we have matched the halo mass and redshift distributions from the simulations to be the same as those in the data. The halo masses of LRGs are estimated using stellar-to-halo masss relations in Section \ref{subsec:halomass} that is applied to the stellar mass distribution of LRGs in Fig.~\ref{f01}. }
\label{f03}
\end{figure*}

The average $y$ profile around 66,479 LRGs is compared to cosmo-OWLS simulations with different AGN feedback models in Fig.~\ref{f03}, where the gray lines show the average $y$ profiles of the simulations. In the comparison, a clear difference between the data and NOCOOL model can be seen, particularly on small angular scales, demonstrating the importance of non-gravitational physics. The incorporation of cooling and heating due to stellar and AGN feedback (AGN 8.0) best matches the data. However, we see a visible trend in that the higher the power of AGN feedback (i.e., increasing the heating temperature, which leads to more violent/bursty feedback), the lower the peak of $y$ profile.  This is due to the fact that the AGN feedback ejects gas from the centre of haloes outwards, lowering the gas density.  We find that the AGN 8.5 and (particularly) AGN 8.7 models yield $y$ profiles that lie below what is observed, at least on scales dominated by the 1-halo term (see below).  A similar result is obtained using the simulations that adopt a \wmap\ cosmology.  This result is consistent with previous studies (e.g., \citealt{Brun2014}), which showed the AGN 8.0 model reproduces a variety of observed gas features in local groups and clusters of galaxies. 

Interestingly, no large difference is seen between the REF and AGN 8.0 models, even though \cite{Brun2015} show their $Y$-$M_{500}$ scalings differ. This can be explained by the fact that the deviation between the REF and AGN 8.0 model only starts to appear below $M_{500} \sim 10^{13.5} \msun$, which roughly corresponds to the average mass of our sample.  The similarity of the observed $y$ signal is due to the similarity of the gas fractions of these two models at the mass scales explored here.  We note, however, that these models differ significantly in their stellar and total baryon fractions and, therefore, comparisons at fixed stellar mass (see Appendix A) show very large differences in the predicted $y$ profile. Thus, the relatively good agreement with the REF model is largely fortuitous and is very much a case of getting the right result for the wrong reason.  Consistent with the findings of many previous studies, we find AGN feedback is required to prevent excess star formation on the scale of groups and clusters.

\section{Halo Model with the UPP}
\label{sec:comp-model}

\begin{figure}
\begin{center}
\includegraphics[width=\linewidth]{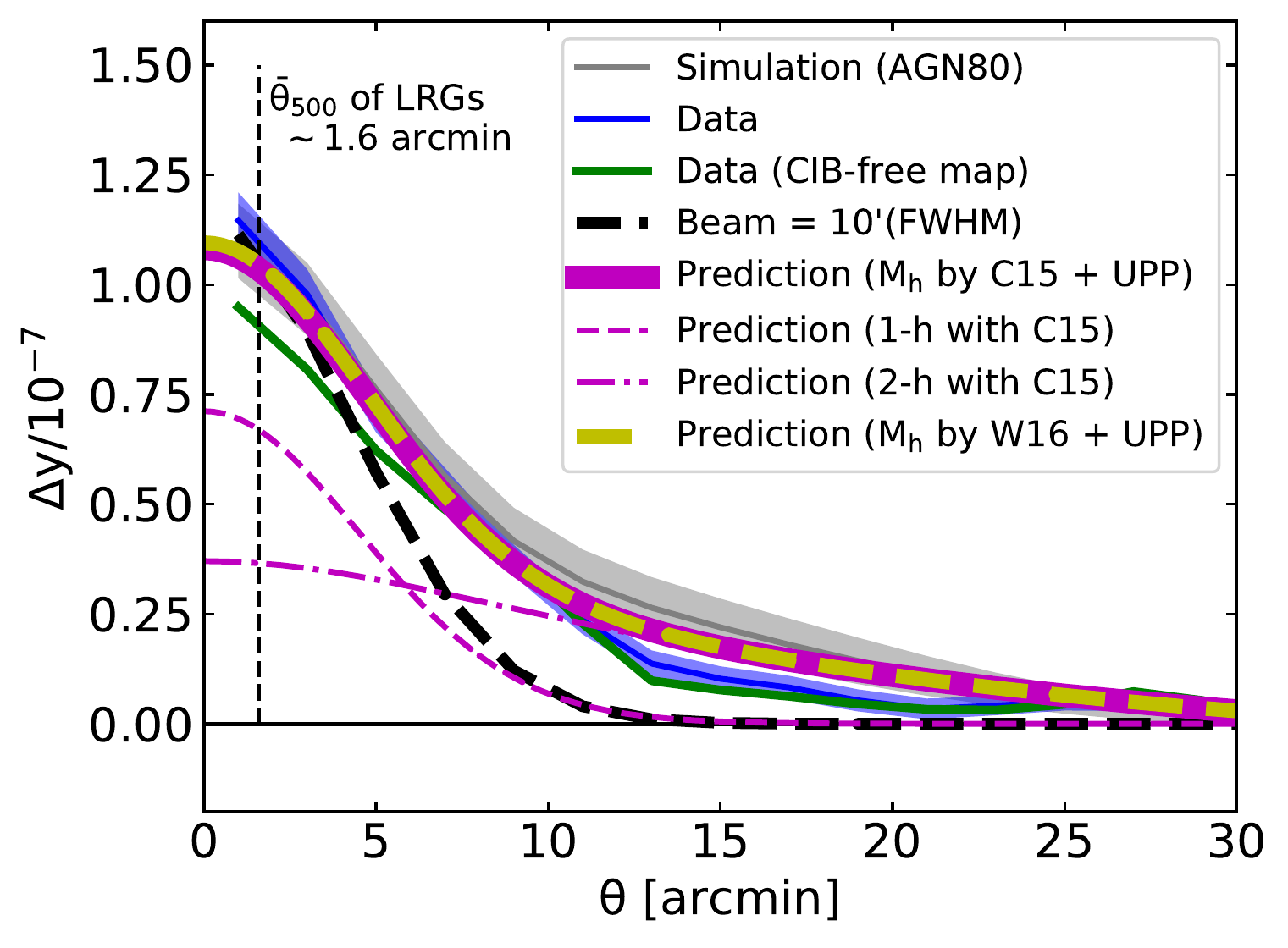}
\caption{The average $y$ profile around 66,479 LRGs (blue) is compared to the predictions using a halo model with the halo mass function and halo bias \citep{Tinker2010} and UPP. The halo masses of the LRGs are estimated using either the SHM relation of C15-SHM (magenta) and W16-SHM (yellow). The one-halo (dashed line in magenta) and two-halo (dash-dotted line in magenta) terms are shown separately for the model prediction using the C15-SHM. The $y$ profile of the simulated central galaxies in the AGN 8.0 simulation is shown in grey.  To show an impact of beam, the average angular size of the LRGs ($\bar{\theta}_{500} \sim 1.6$ arcmin) is shown in vertical black dashed line. Note that the AGN 8.0 simulation is customized to a larger field of view of $\mathrm{10^{\circ} \times 10^{\circ} \, [deg^2]}$ but a limited redshift of $z<1$ in this figure to improve the number of objects as well as background estimates.}
\label{f05}
\end{center}
\end{figure}

Using the estimated halo masses of LRGs, we can calculate the average $y$ profile around LRG haloes using the halo model and UPP via the procedure described in Section \ref{sec:gy-cross}. The model $y$ profiles for two different halo mass estimates are shown in Fig.~\ref{f05} as well as the $y$ profile around the LRGs and the one from the AGN 8.0 simulation. Note that in this analysis we use lightcones from the AGN 8.0 simulation with a larger field-of-view of $\mathrm{10 \times 10 \, deg^2}$ but limited to $z<1$. We do this to improve the number of objects as well as background estimates. We choose the AGN 8.0 simulation because it shows the best agreement with the $y$ profile around the LRGs. 

The predictions from C15-SHM + UPP (magenta) and W16-SHM + UPP (yellow), with the clustering of haloes via a two-halo term properly accounted for (for example, dash-dotted line in magenta), agree well with the observed $y$ profile around the LRGs.  Naively, this is a somewhat surprising result, as \cite{Brun2015} previously showed that the AGN 8.0 simulation predicts a pressure distribution that differs significantly from the UPP at these mass scales.  Yet, the AGN 8.0 model also reproduces our observed stacked profile quite well. 

As discussed in Section \ref{subsec:comp-sim}, \cite{Brun2015} show that the deviation from a power-law relation in the AGN 8.0 simulation begins to appear below $M_{500} \sim 10^{13.5} \msun$, which corresponds roughly to the average mass of our samples.  It implies that stronger deviations from the UPP would be seen in lower-mass haloes than explored here. In addition, the impact of finite resolution is not negligible in our analysis. In particular, the Planck tSZ maps has a FWHM of 10 arcmin.  By comparison, the mean angular size, $\theta_{500}$, of the LRGs is 1.6 arcmin, shown in vertical black dashed line in Fig.~\ref{f05}. Beam smoothing therefore prevents us from placing strong constraints on the tSZ distribution on the scales where the UPP and the simulations differ significantly. Stacked profiles derived from higher-resolution tSZ maps (such as those from ACT or SPT, which have FWHM of order an arcminute) would be very helpful in this regard.

Interestingly, a comparison of the contributions of the one-halo (dashed line in magenta) and two-halo terms (dash-dotted line in magenta) in Fig.~\ref{f05} demonstrates that the two-halo term dominates on scales larger than $\sim6$ arcmin (see also \citealt{Hill2018}).  Given the angular diameter of $\theta_{500}$ noted above (note that $\theta_{200} \approx 2.5$ arcmin), we find that the two-halo term begins to dominate over the one-halo term at approximately $4 r_{500}$ or, roughly, 2 virial radii.  This is what is expected if the halo mass estimates of the LRGs are reliable.

Finally, We estimate the significance of our measured $y$ profile to null hypothesis by measuring the signal-to-noise ratio (SNR). The SNR can be defined as $\sqrt{\Delta \chi^2} = \sqrt{\chi^2_{null} - \chi^2_{bm}}$, where $\chi^2_{null}$ and $\chi^2_{bm}$ refer to the $\chi^2$ statistics applied to the null hypothesis and our halo model prediction using C15-SHM + UPP (magenta in Fig.~\ref{f05}) respectively. They were computed using the covariance matrix accounting for the correlation between different radial bins. The SNR is estimated to be $\sim$17.9.

\section{Further tests for systematics}
\label{sec:systemaics}

To gauge the reliability of our results and conclusions, we have performed a few additional tests.  In particular, we have examined the potential impact of halo mis-centring on the recovered stacked $y$ profile, as well as the potential impact of contamination by the cosmic infrared background (CIB).

If LRGs do not reside at the centres of their host haloes, this will have the effect of artificially lowering our measured $y$ profile. \cite{Reid2009} estimated 89\% of LRGs are central from correlation studies, while \cite{Hoshino2015} found it is only 73\% at a halo mass of $10^{14.5} \msun$. We test for this so-called `mis-centring' effect using the cosmo-OWLS simulations. In the simulations, we artificially shift 27\% (worse case above) of simulated haloes used in Fig.~\ref{f05} by 1 Mpc away from their original positions. Note that 1 Mpc corresponds to $\approx$ 3.6 arcmin at the mean redshift of our LRG sample. We find that the effect of doing this on our stacked $y$ profile is only $\sim$5\% percent and therefore not significant. This is likely due to the coarse angular resolution of the \planck $y$ map. 

Aside from mis-centring, we have also explore the potential impact of contamination of the \planck\ $y$ map due to CIB (e.g., \citealt{Planck2016-XXIII}).  We refer to the study of \cite{Yan2018} who have estimated the CIB contamination of the $Planck$ $y$ maps.  Specifically, they subtracted the \planck\ CIB maps \citep{Planck2016IR-CIB} from the \planck\ intensity maps and reconstructed the CIB-free tSZ map.  We have repeated this procedure and compare the average $y$ profiles of LRGs before and after the CIB subtraction. We find that the amplitude at the locations of our LRG samples is approximately 20\% lower with the `CIB-free' tSZ map , which is shown in green line of Fig.~\ref{f05}. Re-comparing to the simulations in Section \ref{sec:comp-sim} and model predictions in Section \ref{sec:comp-model}, this may suggest that somewhat more aggressive AGN feedback (relative to the AGN 8.0 model) is required to match the data.  Furthermore, it may provide evidence for a small deviation from the predictions of the UPP.  However, since the tSZ effect and CIB signals are known to be correlated \citep{Addison2012, Planck2016-XXIII}, the subtraction of the CIB maps will likely have removed some of the tSZ signal itself and the actual CIB contamination would therefore be less than the estimated above. We therefore cannot make definitive statements about the required feedback energetics or the presence of small deviations from the UPP.  We stress, however, that our general conclusion (i.e., that efficient feedback from AGN is required to reproduce the observed signal) is insensitive to uncertainties in the treatment of CIB contamination.

\section{Discussion}
\label{sec:discussion}

This study was partially motivated by an apparently contradictory result between \cite[P13]{Planck2013IR-XI} and \cite[A15]{Anderson2015} on the state of hot gas in galaxy group/clusters through scaling relations. A self-similar scaling relation between halo electron pressure and halo mass is valid under the assumption that the galaxy-formation process is dominated by gravity; any deviation from this relation points to the presence of more complex processes such as baryonic feedback effects. Using the Locally Brightest Galaxies (LBGs) in SDSS DR7, P13 find the self-similar scaling relation in $Y$-$M_{\rm h}$, therefore implying that gravity is dominant even in low-mass haloes and that they incorporate the mean cosmic fraction of baryons as seen in more massive haloes, with the assumption that the gas in low-mass haloes is in a virialized state. On the other hand, A15 finds a steeper scaling than the self-similar scaling relation in $L_{\rm X}$-$M_{\rm h}$, suggesting the importance of non-gravitational heating such as AGN feedback. Numerous X-ray studies of galaxy groups also find a deficit of baryons inside low-mass haloes compared to the cosmological mean (e.g., \citealt{Gastaldello2007, Pratt2009, Sun2009, Gonzalez2013}). These results can be reconciled by the idea that low-mass haloes may contain the cosmic fraction of baryons, just like galaxy clusters, but with a density profile of gas that is less centrally concentrated. In other words, that groups and clusters do reach the same cosmic fraction but only on scales larger than typically probed with X-ray observations (but which can be probed by the tSZ effect). \cite{Brun2015} tested the \planck result using the cosmo-OWLS simulations and showed that the tSZ flux within $R_{500}$ is highly sensitive to the assumed pressure distribution of the gas and, given the pressure profiles from the AGN 8.0 model, showed that the self-similar model would not be valid in low-mass haloes, at least on small scales. 

We find that the measured $y$ profile around LRGs agrees best with the profile measured from the AGN 8.0 simulation, which was shown in previous studies to also provide a good match to the observed X-ray scaling relations of groups and clusters.  A model that neglects non-gravitational physics altogether (i.e., NOCOOL), produces observed $y$ profiles in excess of what is observed, while models that adopt very violent/bursty AGN feedback lower the predicted $y$ profile below that observed on small scales.
We also demonstrate that the measured $y$ profile around LRGs agree with the predictions using the UPP given the SHM relation from C15-SHM \citep{Coupon2015} or W16-SHM \citep{Wang2016}, estimated by gravitational lensing measurements. This implies that the UPP, estimated for galaxy clusters in the mass range of $10^{14.4} - 10^{15.3} \msun$, can also be applied to low-mass systems down to $\sim 10^{13.5} \msun$.  However, we cannot rule out small deviations from the UPP due to uncertainties related to CIB contamination for the $Planck$ $y$ maps (see Section \ref{sec:systemaics}).

Interestingly, the AGN 8.0 simulation predicts more extended pressure profiles around low-mass haloes than the UPP (see \citealt{Brun2015}) and also reproduces the observed $y$ profile.  This apparent inconsistency is explained by the fact that the deviations from the self-similar model in \citet{Brun2015} are mainly confined to halo masses below $M_{500} \sim 10^{13.5} \msun$, which roughly corresponds to the average mass of our sample.  Furthermore, the impact of coarse angular resolution of the \planck $y$ map is not negligible in our analysis: the UPP and AGN 8.0 pressure distributions only differ significantly on scales of $r \la r_{500}$, which are well within the beam. Data from higher-resolution tSZ maps (such as those from ACT or SPT, which have FWHM of order an arcminute) would be important in this regard.

\section{Conclusion}
\label{sec:conclusion}
In this paper we have presented a stacking analysis of the $y$ signal measured by \planck around SDSS DR7 LRGs, which are considered to be mostly central galaxies in dark matter haloes. We construct the average $y$ profile centred on the LRGs and study the thermodynamic state of the gas in groups and low-mass clusters. The major results of our analysis are summarized as follows: 

\begin{itemize}
\item We detect a significant tSZ signal out to $\sim$ 30 arcmins well beyond the extent of the 10 arcmin beam of the \planck $y$ map.
\item We compare the average $y$ profile around LRGs with the predictions from the cosmo-OWLS suite of cosmological hydrodynamical simulations. This comparison agrees best with simulations that include AGN feedback (AGN 8.0), but not with simulations that do not include non-gravitational physics (NOCOOL) or with simulations with very violet AGN feedback (AGN 8.5, AGN 8.7). This is consistent with other studies showing that the AGN 8.0 model reproduces a variety of observed gas features in optical and X-ray data (e.g., \citealt{Brun2014}). The data also agree with the REF model that includes cooling and heating due to stellar feedback, but no AGN feedback. This can be explained by \cite{Brun2015} showing that the deviation between the REF and AGN 8.0 model starts to appear below $M_{500} \sim 10^{13.5} \msun$, which almost corresponds to the average mass of our samples.  We note, however, that models that neglect AGN feedback lead to excessive star formation and overcooled massive galaxies.  Consequently, an analysis of the stacked $y$ profiles in bins of stellar mass (see Appendix A), clearly rules out the REF model.
\item The average $y$ profile around the LRGs is also compared with a prediction using the halo model with a UPP. The predicted $y$ profile is consistent with the data, but only if we account for the two-halo clustering term in the model, and if we assume the stellar-halo mass relation from either C15-SHM or W16-SHM, which are estimated using gravitational lensing measurements. This may imply that the UPP,  estimated for massive galaxy clusters in the mass range of $10^{14.4} - 10^{15.3} \msun$, can be applicable even in low-mass haloes down to $\sim 10^{13.5} \msun$. 
\end{itemize}

In our analysis, the dominance of the two-halo term in low-mass systems is partially due to the coarse angular resolution of the \planck $y$ map. We emphasize that more precise measurements with a better angular resolution and sensitivity such as ACTPol \citep{Niemack2010} and SPTpol \citep{Austermann2012} will shed further light on the issue and help to clarify the impact of AGN feedback on the formation and evolution of galaxies. 

\section*{Acknowledgement}

The authors thank Amandine Le Brun and Joop Schaye for their contributions to the cosmo-OWLS simulations.

This research is funded by (1) Canada's NSERC and CIFAR. This research has been also supported by funds from (2) the European Research Council (ERC) under the Horizon 2020 research and innovation programme grant agreement of the European Union: ERC-2015-AdG 695561 (ByoPiC, https://byopic.eu), (3) the ERC under the European Union's Horizon 2020 research and innovation programme (grant agreement No 769130), (4) AJM and TT acknowledge support from the Horizon 2020 research and innovation programme of the European Union under the Marie Sk\l{}odowska-Curie grant agreements No.~702971 and No.~797794, (5) AJM acknowledges support from a CITA National Fellowship and (6) Y.Z.M. acknowledges the support by National Research Foundation of South Africa (No. 105925, 109577, 110984) and UKZN Hippos cluster.

The data is based on observations obtained with Planck (http://www.esa.int/Planck), an ESA science mission with instruments and contributions directly funded by ESA Member States, NASA, and Canada, and observations obtained with SDSS-I and SDSS-II, managed by the Astrophysical Research Consortium for the Participating Institutions (http://www.sdss.org/) funded by the Alfred P. Sloan Foundation, the Participating Institutions, the National Science Foundation, the U.S. Department of Energy, the National Aeronautics and Space Administration, the Japanese Monbukagakusho, the Max Planck Society, and the Higher Education Funding Council for England. 



\footnotesize{
\setlength{\bibhang}{2.0em}
\setlength\labelwidth{0.0em}
\bibliographystyle{mnras}
\bibliography{LRGsingle}
}





\appendix
\section{Comparison with simulations with stellar mass}
We compare the average $y$ profile to simulations using stellar masses, instead of halo masses. To identify simulated LRGs, we select simulated central galaxies with the same stellar mass and redshift ranges as in the real data. The average stacked $y$ profile in each stellar mass and redshift bin is then constructed from the simulated light cones. The stacks are then combined, weighted by the total number of LRGs as described in Section \ref{subsec:comp-sim}. 

\begin{figure*}
\begin{center}
\begin{minipage}{0.49\linewidth}
\includegraphics[width=\linewidth]{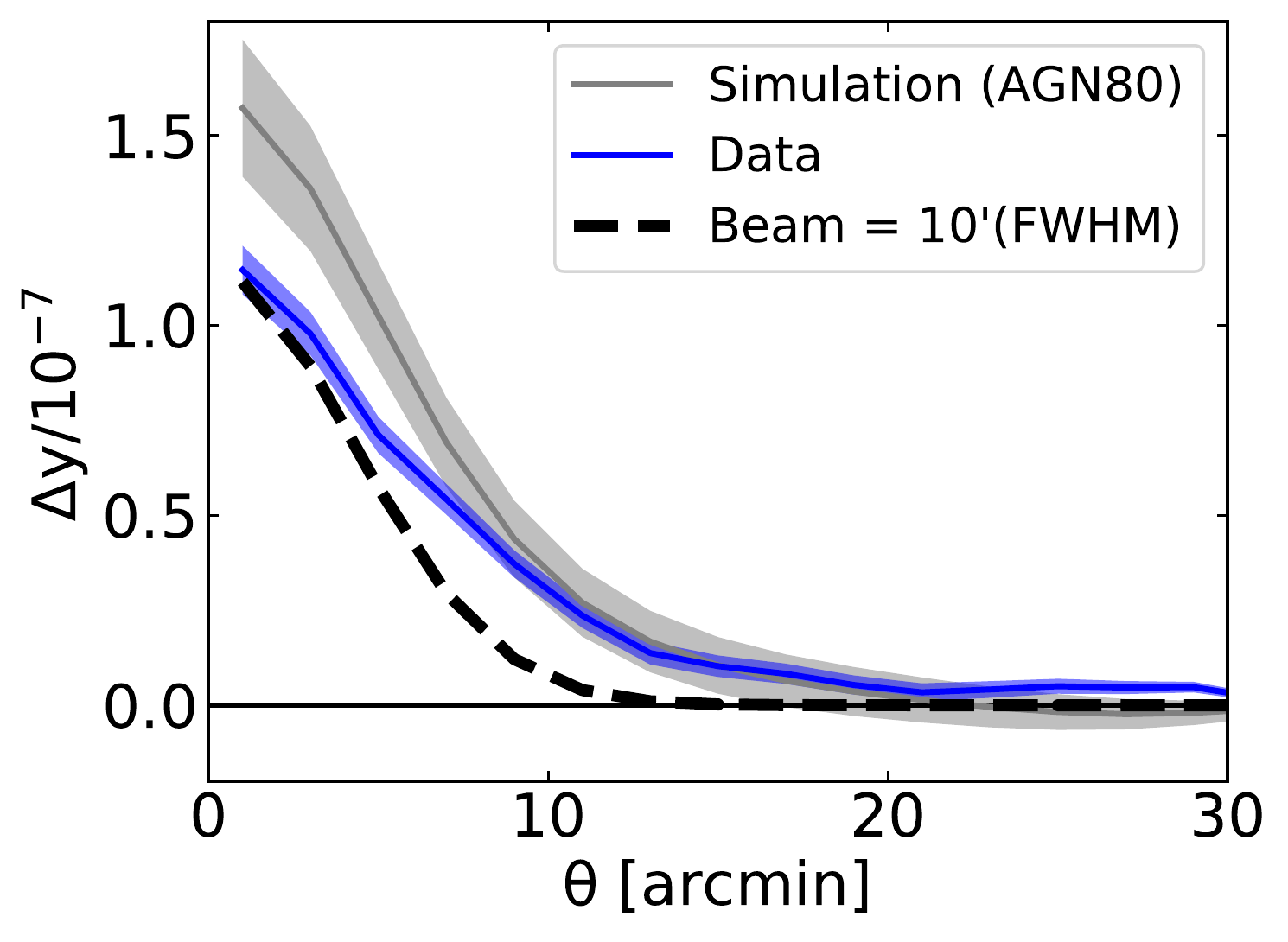}
\end{minipage}
\begin{minipage}{0.49\linewidth}
\includegraphics[width=\linewidth]{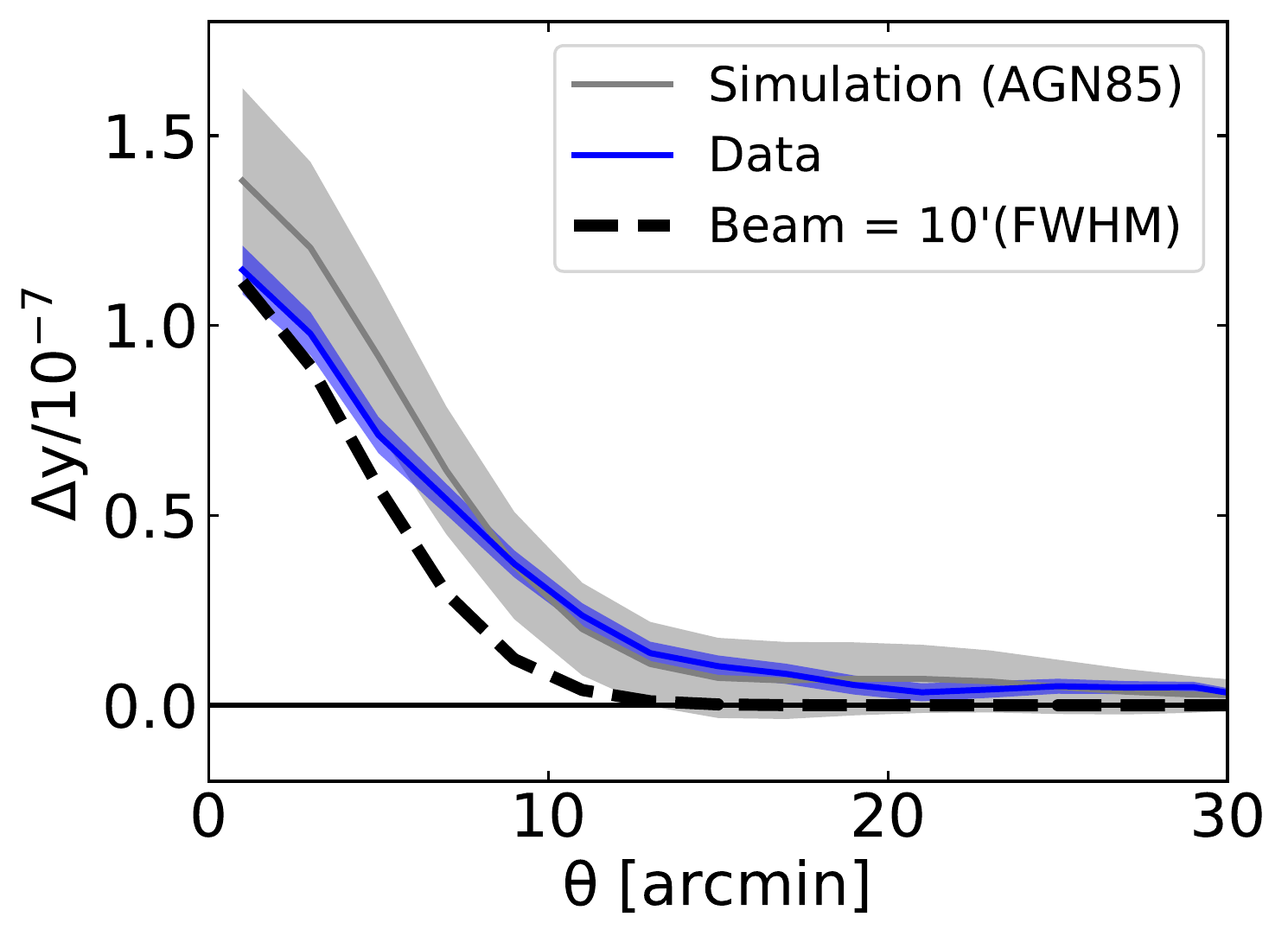}
\end{minipage}
\begin{minipage}{0.49\linewidth}
\includegraphics[width=\linewidth]{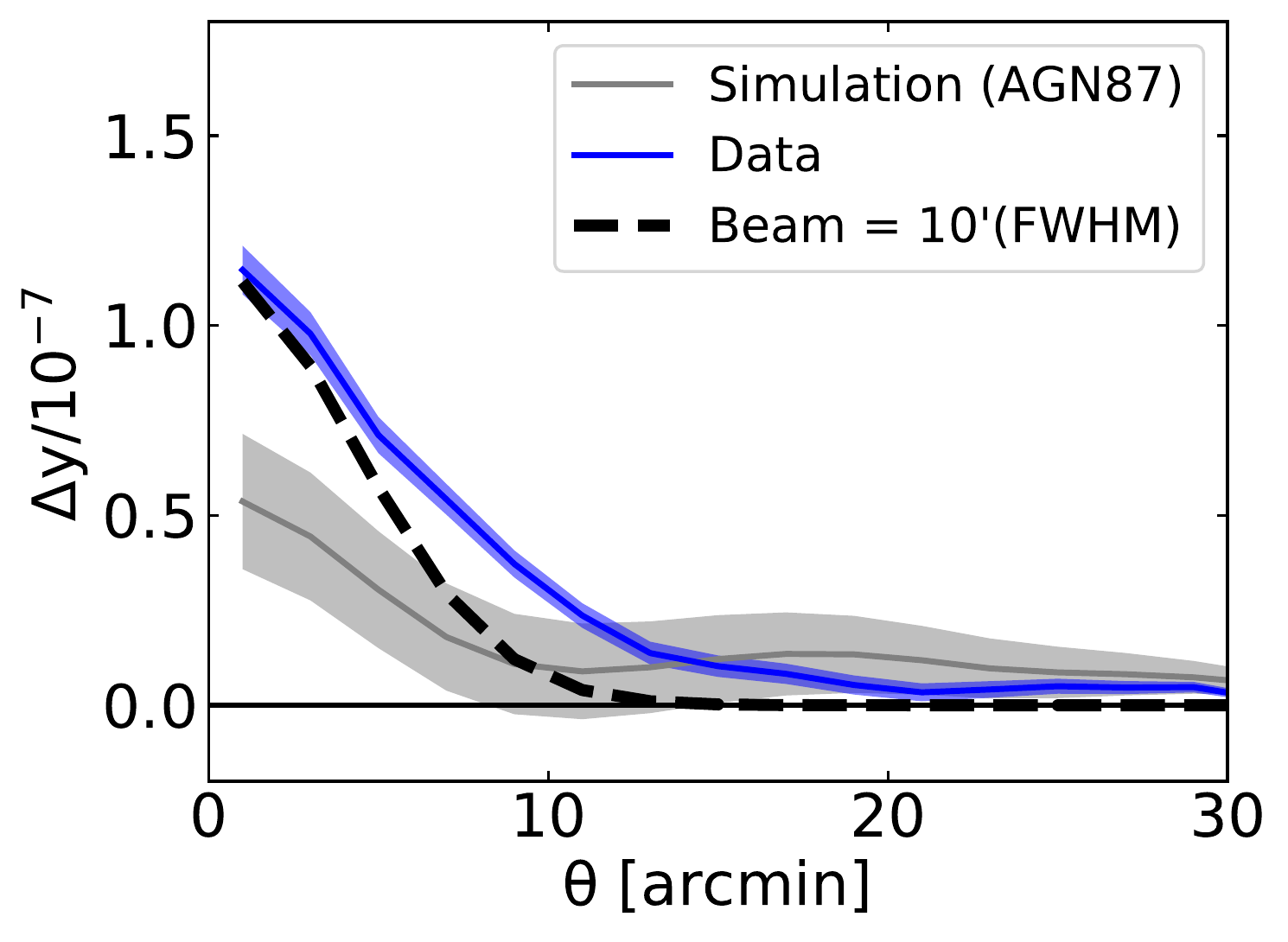}
\end{minipage}
\begin{minipage}{0.49\linewidth}
\includegraphics[width=\linewidth]{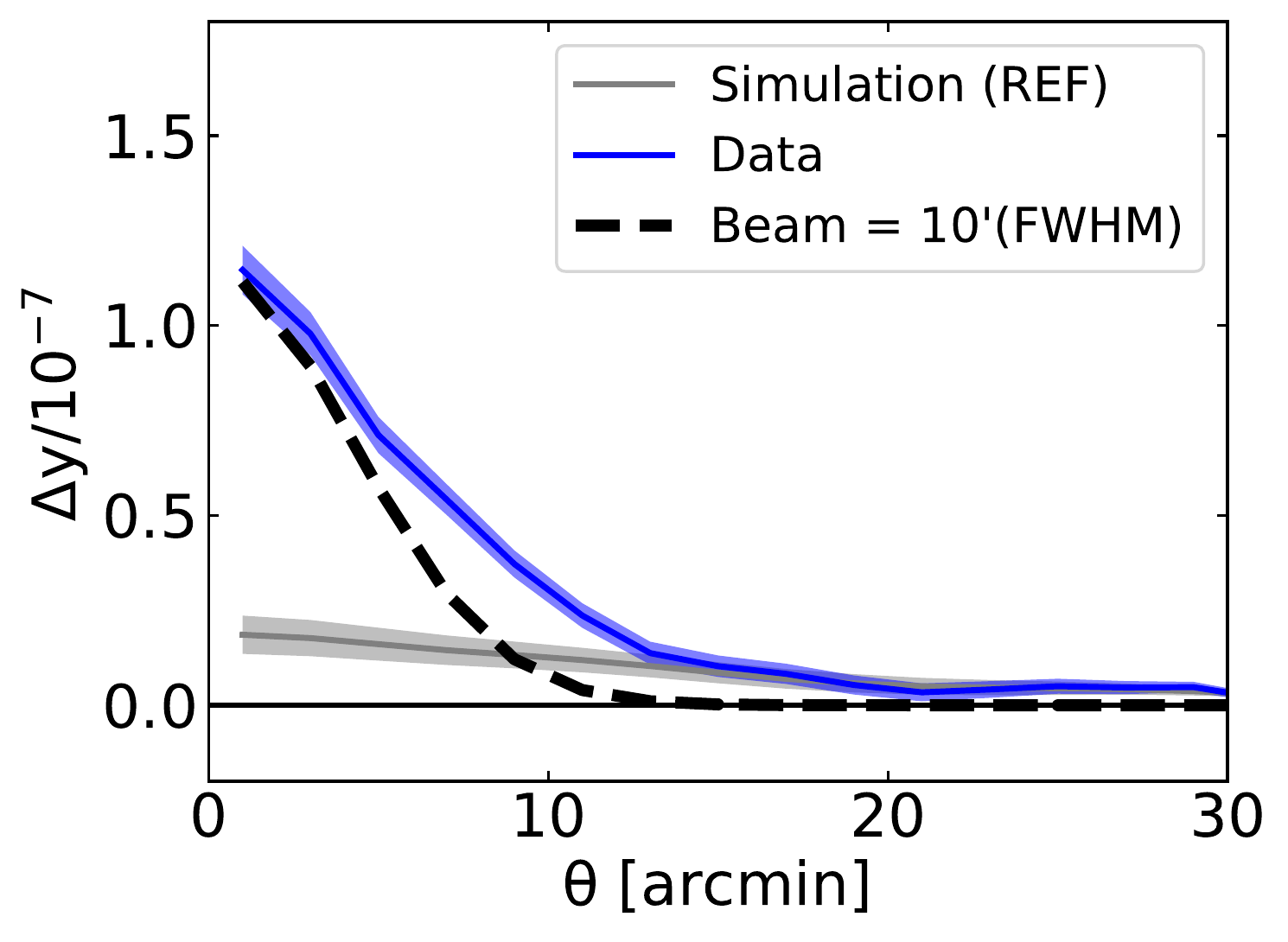}
\end{minipage}
\end{center}
\caption{The average $y$ profile around LRGs (blue) is compared to the $y$ profiles of the simulated central galaxies (gray) in different AGN feedback models respectively. In each case we have matched the stellar mass and redshift distributions from the simulations to be the same as those in the data. {\it Top left}: AGN 8.0 model, {\it Top right}: AGN 8.5 model, {\it Bottom left}: AGN 8.7 model, and {\it Bottom right}: REF model.  }
\label{a01}
\end{figure*}

The average $y$ profile around 66,479 LRGs is compared to cosmo-OWLS simulations with different AGN feedback models in Fig.~\ref{a01}, where the gray lines show the average $y$ profiles of the simulations.  We exclude the NOCOOL model from this comparison, since it does not form galaxies (i.e., no stellar masses).
In the comparison, a clear difference between the data and REF model can be seen. 

In general, energy released from the centre of a halo heats cluster gas, this in turn prevents cooling and thus the star formation around the central region. Therefore, if we consider haloes of the same total mass, the stellar mass of the central galaxy is decreased as the power of the central AGN is increased. Since we select central galaxies based on stellar mass, lower-mass haloes are selected in the REF model compared to the models that include AGN feedback. This is apparent as the lower central peak value of the simulated $y$ profiles in the REF model compared to the AGN models. We also see a visible trend in that the higher the power of AGN feedback, the lower the peak of $y$ profile. This is due to the fact that the AGN feedback ejects gas from the centre of haloes outward and the overall gas density is lowered.  Note that the three AGN models have approximately the same galaxy stellar mass function \citep{McCarthy2017}, so differences in the stacked $y$ profiles indicate real differences in the pressure distribution of the hot gas (as demonstrated at fixed halo in the main text).

As a result of this comparison, we can strongly rule out the REF model. Interestingly, in bins of stellar mass, we find that the AGN 8.5 model reproduces the observed $y$ profile the best, whereas the comparison at fixed halo mass suggested a somewhat better fit by the AGN 8.0 model (modulo possible CIB contamination, which would affect both comparisons in the same way).  This discrepancy may be caused by the fact that the comparison of the simulations to the data is not entirely straightforward. In particular, the methods for estimating stellar masses are different. The stellar mass in the data is estimated by fitting the five-band SDSS photometry to $\sim$400 spectral templates and adopting a particular stellar population synthesis package and an assumed stellar initial mass function.  On the other hand, the stellar mass for the simulated galaxies is estimated by simply summing the masses of star particles within 30 kpc around central galaxy.  In terms of observational systematics alone, the typical uncertainty (excluding uncertainties in the stellar initial mass function) is $\sim0.25$ dex in stellar mass (e.g., \citealt{Behroozi2010}).  While this may explain the difference in the preference of somewhat different AGN feedback models, note that it cannot reconcile the REF model with the data, as the REF model predicts stellar masses that at least 0.5 dex too large (see figure 1 of \citealt{McCarthy2017}). 

\bsp	
\label{lastpage}
\end{document}